\documentclass[manuscript]{aastex} % Default class - double spaced, single column
\usepackage{epstopdf}
\slugcomment{Accepted for publication in The Astrophysical Journal, February 2015}

\shorttitle{Solar Flare Abundances}
\shortauthors{Dennis et al.}

\begin{document}

\title{Solar Flare Element Abundances from the Solar Assembly for X-rays (SAX) on MESSENGER}

\author{Brian R. Dennis\altaffilmark{1}}
\email{brian.r.dennis@nasa.gov}

\author{Kenneth J. H. Phillips\altaffilmark{2}, Richard A. Schwartz\altaffilmark{1, 3}, Anne K. Tolbert\altaffilmark{1, 3}, Richard D. Starr\altaffilmark{3, 4}, and Larry R. Nittler\altaffilmark{5}}

\altaffiltext{1}{Solar Physics Laboratory (Code 671), Heliophysics Science Division, NASA Goddard Space Flight Center, Greenbelt, MD 20771, USA}
\altaffiltext{2}{ Earth Sciences Department, Natural History Museum, London SW7 5BD, UK}
\altaffiltext{3}{The Catholic University of America, 620 Michigan Ave.~NE, Washington, DC 20064, USA}
\altaffiltext{4}{Astrochemistry Laboratory (Code 691), Solar System Exploration Division, NASA Goddard Space Flight Center, Greenbelt, MD 20771, USA}
\altaffiltext{5}{Department of Terrestrial Magnetism, Carnegie Institution of Washington, 5241 Broad Branch Road NW, Washington, DC 20015, USA}

\author{}

\begin{abstract}

X-ray spectra in the range $1.5-8.5$~keV have been analyzed for 526 large flares detected with the Solar Assembly for X-rays (SAX) on the Mercury {\em MESSENGER} spacecraft between 2007 and 2013. For each flare, the temperature and emission measure of the emitting plasma were determined from the spectrum of the continuum.  In addition, with the SAX energy resolution of 0.6 keV (FWHM) at 6~keV, the intensities of the clearly resolved Fe-line complex at 6.7~keV and the Ca-line complex at 3.9~keV were determined, along with those of unresolved line complexes from S, Si, and Ar at lower energies.  Comparisons of these line intensities with theoretical spectra allow the abundances of these elements relative to hydrogen to be derived, with uncertainties due to instrument calibration and the unknown temperature distribution of the emitting plasma. While significant deviations are found for the abundances of Fe and Ca from flare to flare, the abundances averaged over all flares are found to be enhanced over photospheric values by factors of $1.66 \pm 0.34$ (Fe), $3.89~\pm~0.76$ (Ca), $1.23~\pm~0.45$ (S), $1.64~\pm~0.66$ (Si), and $2.48~\pm~0.90$ (Ar).  These factors differ from previous reported values for Fe and Si at least.  They suggest a more complex relation of abundance enhancement with the first ionization potential (FIP) of the element than previously considered, with the possibility that fractionation occurs in flares for elements with a FIP of less than $\sim$7~eV rather than $\sim10$~eV.

\end{abstract}

\keywords{Sun: abundances --- Sun: corona --- Sun: flares --- Sun: X-rays, gamma rays  --- line: identification}

%%%%%%%%%%%%%%%%%%%%%%%%%%%%%%%%%%%%%%%%%%%%%%%%%%%%%%%%%%%%%%%%%%%%%%%%%%%%%%%%%%%%%%%%%%%%%%%%%%%%%%%%%%%%%%%%%%%%%%%

\section{INTRODUCTION}
\label{intro}

The Solar Assembly for X-rays (SAX) is part of the X-ray Spectrometer (XRS: \cite{sch07}) on the {\em MESSENGER} (MErcury Surface, Space ENvironment, GEochemistry, and Ranging) mission \citep{2001P&SS...49.1445S, 2001P&SS...49.1467G, 2001P&SS...49.1481S}. Its primary role is to measure the spectrum of solar soft X-rays that are the main exciters of fluorescence radiation from Mercury's surface.  The spectrum of this fluorescence radiation is measured, in turn, with three gas proportional counters of XRS to provide information on the chemical composition of the surface \citep{nit11}. SAX consists of a cooled, 500~$\mu$m thick Si--PIN solid-state detector covering the energy range from $\sim$1 to $\sim$10~keV with an energy resolution of 598~eV at 5.9 keV. Its pinhole aperture has an area of only 0.03~mm$^2$ so that the detector count rates are limited to manageable levels during all but the most intense flares. The detector assembly is mounted on the {\em MESSENGER} sunshade and views the Sun directly through three Be foils (each 25$\mu$m thick) to reduce the solar thermal input to the detector and  attenuate the flux of the lowest energy X-rays.

Since the launch of {\em MESSENGER} on 2004 August~3 and capture into orbit around Mercury in 2011, SAX has observed hundreds of solar flares and obtained soft X-ray spectra with a cadence generally of 5 minutes. The emission in the 1-10~keV energy range is dominated by free--free and free--bound thermal continuum radiation but also includes spectral line complexes that can be resolved with the SAX spectral resolution. These include complexes of lines due to highly ionized Fe (peaking at $\sim 6.7$~keV and $\sim 8$~keV) and Ca (peaking at $\sim 3.9$~keV). Estimates of the abundances of Fe and Ca can be made from spectral fits to the measured count-rate spectra with a precision depending primarily on the accuracy of the assumed temperature distribution. Forms for the distribution (the differential emission measure) which are a power-law or exponential in temperature were considered, but most of the analysis was done with simpler forms described by a single temperature (isothermal) or two temperatures. At energies of $\lesssim 3.5$~keV, contributions to the emission are also made by lines of S, Si, Ar, Mg, and Fe ions but these are not resolved with SAX.  Consequently, estimates of the S, Si, and Ar abundances can also be made but with greater uncertainty because of their smaller contributions to the total emission. Mg makes too small a contribution to the spectrum and at too low an energy for a meaningful abundance determination to be made from SAX spectra. This paper is primarily concerned with deriving estimates of the abundances of these elements for flares observed with SAX and comparison with those from other instruments and with theory.

Element abundances in coronal plasmas such as flares are widely considered to be different from those in the photosphere by amounts that depend on the first ionization potential (FIP) of the element -- the so-called FIP effect -- but there is still debate about the magnitude of the difference for each element or the so-called FIP bias, equal to the ratio of the measured element abundance to the abundance in the photosphere. \cite{fel92} give values of element abundances in the corona that are as much as a factor of 4 higher than photospheric abundances (a FIP bias of $\sim 4$) for elements with relatively low FIP (taken to be $\lesssim 10$~eV) but are approximately equal to photospheric abundances (FIP bias of 1) for high-FIP ($\gtrsim 10$~eV) elements. The later work of \cite{fel00} gives reduced abundances of some low-FIP elements, notably Mg, Si, and Ca. The \cite{fel92} values are the basis for the ``coronal" abundance set in the {\sc chianti} database and software package used extensively here \citep{der97,lan13}. \cite{flu99} propose coronal abundances (the ``hybrid" abundance set in {\sc chianti}) that are enhanced over the photospheric abundance estimates available at that time for low-FIP elements by factors of $\sim$2 but reduced by factors of up to $\sim$2 for high-FIP elements. \cite{sch12} have given recommended abundances for the solar corona but this is an amalgam of results from coronal spectroscopy, the solar wind, and solar energetic particles so it may not be appropriate for solar flares. The most recent photospheric abundance estimates \citep{asp09,caf11} have resulted in some minor revisions of elements considered here (Fe, Ca, S, and Si) compared with previous work (although there are substantial reductions in the C, N, and O abundances). The photospheric abundances used here to evaluate FIP bias values given in Table \ref{tab:JuneflareFIP&FIPbias} are from \cite{asp09}.
% except for Ar which is taken from with \cite{lod08} based on solar proxies.

More refined estimates of coronal abundances of S, Si, Ar, and K have been made from flare and non-flare observations of X-ray lines by the RESIK crystal spectrometer on {\it Coronas-F} \citep[see][and references therein]{bsyl14} and of Fe from {\it RHESSI} observations of the line complexes at 6.7~keV and 8~keV \citep{phi12}. Compared with the photospheric abundances of \cite{asp09}, they indicate FIP biases which are listed in Table \ref{tab:JuneflareFIP&FIPbias}.
There is some disagreement between these estimates and those of \cite{flu99} (for S) and \cite{fel00} (for Fe). The RESIK and {\em RHESSI} spectra thus suggest that the dependence on FIP is not as simple as that given originally by \cite{fel92}. This is also indicated by some more recent observational results which we will discuss later.

Theoretical models to explain the FIP effect generally use the fact that low-FIP elements exist in a partially or wholly ionized form in the photosphere or chromosphere but high-FIP elements are neutral. Earlier models, summarized by \cite{hen98}, include mechanisms like diffusion, with magnetic fields playing an active or passive role. In the model developed by \cite{lam12} (and references therein), FIP-dependent abundance enhancements occur as the result of ponderomotive forces associated with Alfv\'{e}n waves excited in the solar corona, and can be calculated for different values of coronal flux loop lengths and wave flux. Thus, comparison with observations is possible in principle.

In this paper, we report on an extensive analysis of {\em MESSENGER} SAX spectra taken during 526 well-observed flares detected between 2007 and 2013, when the spacecraft was first in its cruise phase to Mercury and then in orbit about Mercury. Abundance estimates for the elements Fe, Ca, S, Si, and Ar are given and their distributions examined. The observations and analysis procedures are described in Section~2 for both a single flare and for all flares in the data set with sufficient intensity to detect emission in the Fe-line complex at 6.7 keV. The resulting abundance values are given in Section~3, and compared in Section~4 with previous results and with the predictions of the \cite{lam12} theory.

%%%%%%%%%%%%%%%%%%%%%%%%%%%%%%%%%%%%%%%%%%%%%%%%%%%%%%%%%%%%%%%%%%%%%%%%%%%%%%%%%%%%%%%%%%%%%%%%%%%%%%%%%%%%%%%%%%%%%%%

% Section 2
\section{OBSERVATIONS AND DATA ANALYSIS}

%Subsection 2.1
\subsection{Flare Selection and Time Coverage}
\label{flareselection}

SAX data were normally collected continuously over 5-minute intervals although on some occasions the instrument was commanded into a high-cadence mode with 20-s or 40-s data-gathering intervals close to periapsis of the orbit about Mercury and during some of the more intense flares. Only those time intervals in which the count rate was $>5\,000$~s$^{-1}$ in the 6.3--7.0~keV energy range were selected for analysis to ensure that the Fe line complex at 6.7~keV was well observed. This resulted in the selection of 656 flares or significant increases in X-ray flux for analysis. Of these, events lasting for only a single 5-minute data-gathering interval were eliminated from this data set leaving a total of 526 flares with at least two contiguous data intervals that were used to determine element abundances.  The results from these flares were used in computing the average abundances listed in Table \ref{tab:table2}.  A full catalog of flares analyzed with best-fit spectral parameters and more extensive plots are available at the following location:
\newline
http://hesperia.gsfc.nasa.gov/messenger/results/plots\_and\_tables/.

The analyzed flares covered the period from 2007 May~28 to 2013 August~19.  They are not uniformly distributed in time for three main reasons: the varying solar activity through the deep minimum in 2009, intermittent detector operations necessitated by the various spacecraft maneuvers and periods of high detector temperatures, and particle contamination when the spacecraft was presumably well connected magnetically to the flaring active region.  As a result, there are 27 flares analyzed up until 2007 December 31 and the remainder after 2010 February 6.  An analysis of eleven flares in 2011 has previously been given by \cite{nit11}.  Four of these satisfy our selection criteria and so are included in our lists of analyzed events. The others were either too small to be included or occurred during a period of high particle contamination.

% Subsection 2.2
\subsection{Analysis of Flare on 2007 June 1}

\begin{figure}
    \includegraphics*[angle = 0,width = 0.5\textwidth] {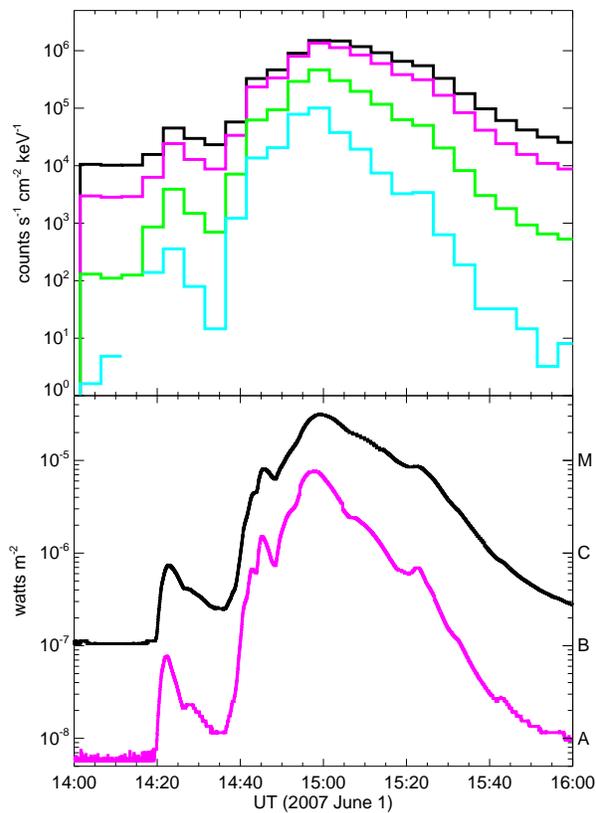}
    \caption{SAX and {\em GOES}~light curves for the M2.8 flare on 2007 June~1. Top panel: SAX light curves plotted with 300~s time bins in four energy bands: $1.3 - 2$~keV (black), $2 - 3.2$~keV (pink), $3.2 - 5.3$~keV (green), $5.3 - 8.5$~keV (cyan). Bottom panel: {\em GOES} X-ray light curves at $1-8$~\AA\ (black) and $0.5-4$~\AA\  (pink) with 3~s time bins.}
    \label{fig:lc_sax_hsi_goes}
\end{figure}

A detailed analysis of an M2.8 flare observed by SAX on 2007 June~1 is first given as an illustration of the methods we used. Figure~\ref{fig:lc_sax_hsi_goes} shows X-ray light curves as recorded by SAX and {\em GOES}. Since the event was from NOAA active region 10960 at S08E78 and Mercury was at a Stonyhurst longitude of 49$^\circ$ East, both instruments would have seen the event equally well. The {\em GOES} light curves show an initial peak at 14:22~UT followed by the main event peaking at 14:59 UT, then a gradual decay to the pre-flare level at about 16:15~UT. As would be expected from their sensitivity to similar energy ranges, the SAX light curves (corrected for the light travel time from {\em MESSENGER} to 1 AU) closely match the {\em GOES} light curves, although with a time resolution of 300~s compared with 3~s for {\em GOES} (2~s after 2009 December 2 with {\em GOES~14} and {\em GOES~15}).

% Figure 1

\begin{figure}
    \includegraphics[angle = 90, width = 0.5\textwidth,] {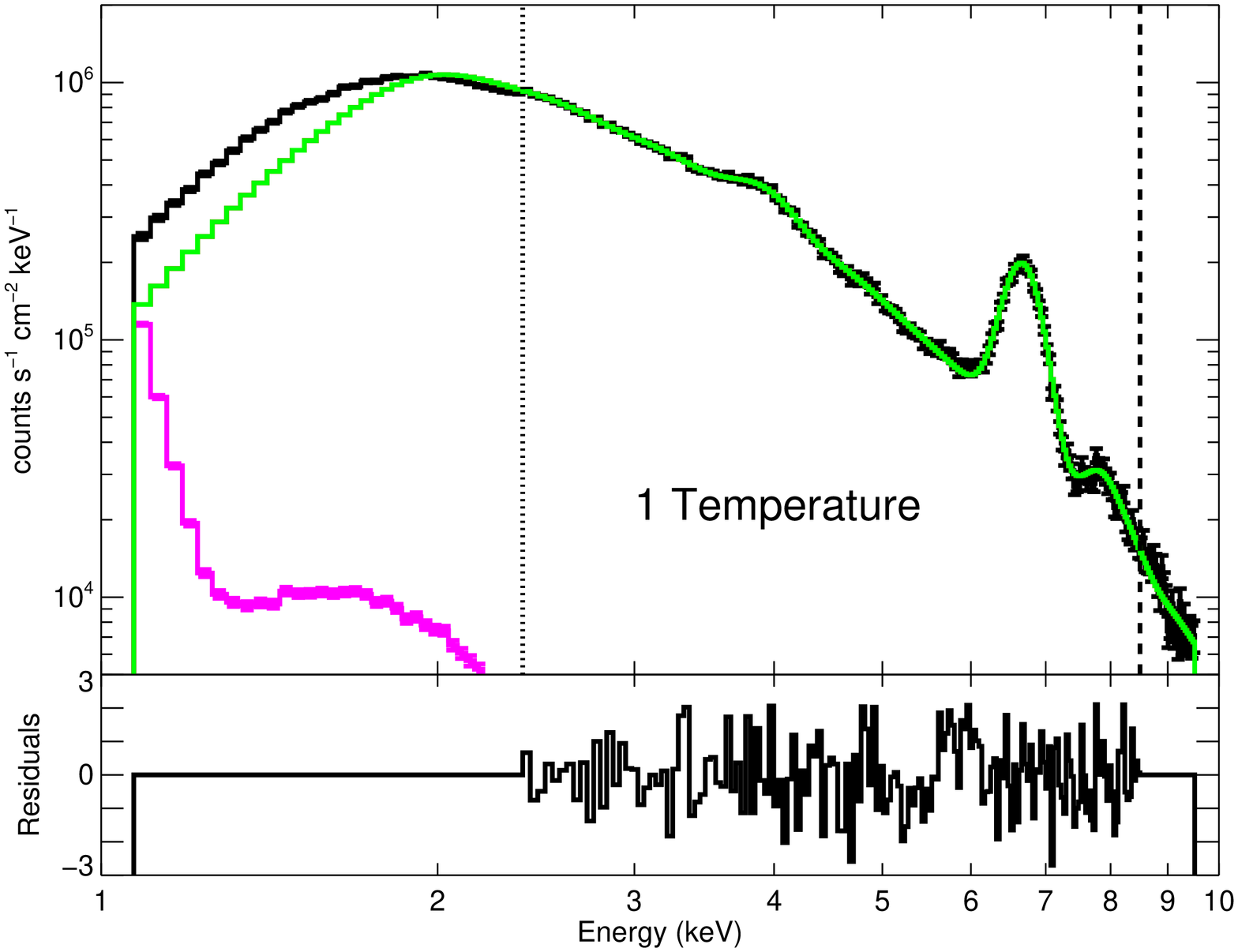}
    \includegraphics[angle = 90, width = 0.5\textwidth,] {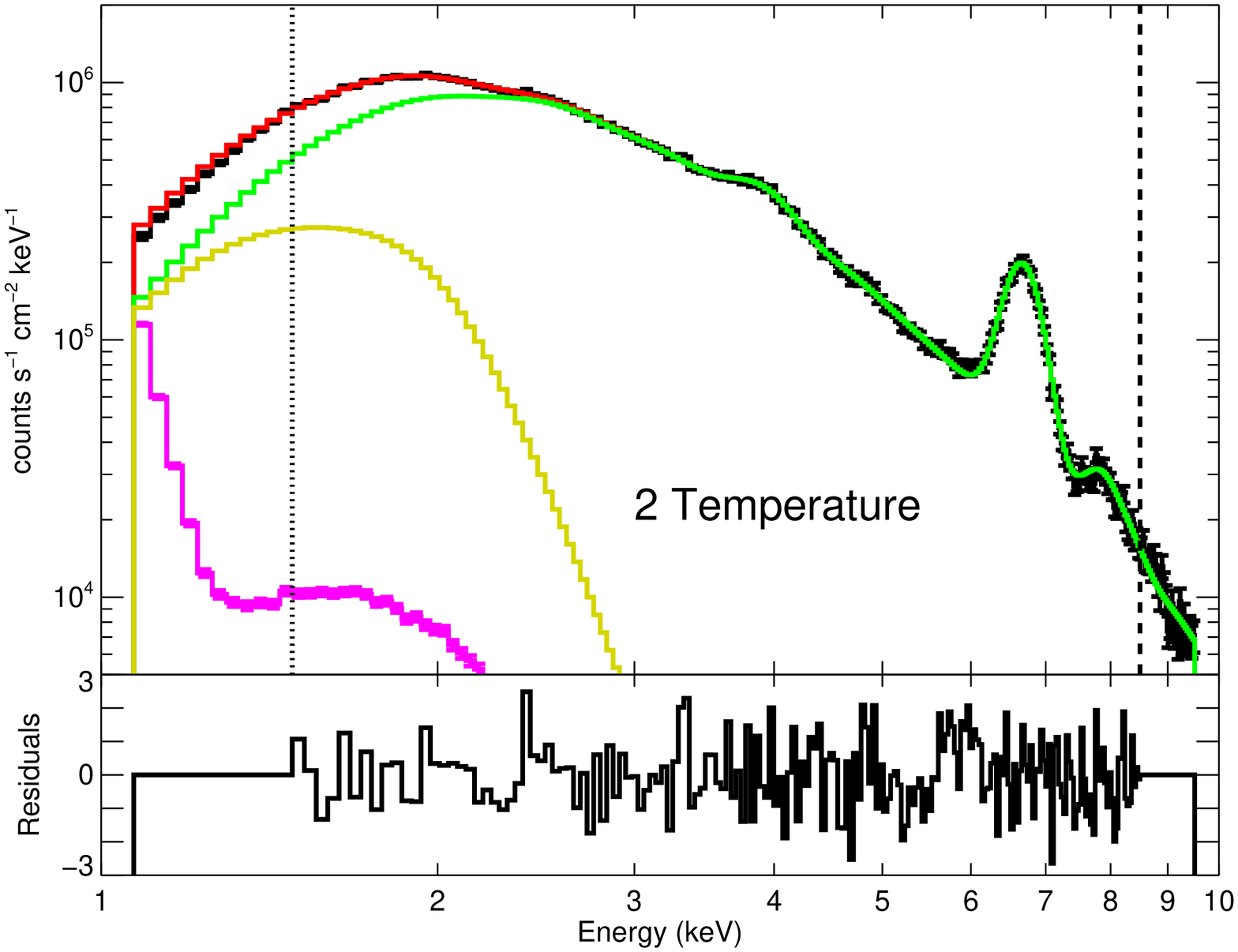}
    \caption{SAX background-subtracted count flux spectra (black histograms) and two model fits for the 14:56:30--15:01:29~UT period at the maximum of the 2007 June~1 flare with normalized residuals (the differences between the measured and best-fit fluxes as a fraction of the statistical $1\sigma$ uncertainties) shown below each spectrum. Left: \mbox{1-T} (isothermal) function (green curve) fitted over the $2.3-8.5$~keV energy range with $T=22$~MK (1.9~keV), EM = $7.1 \times 10^{48}$~cm$^{-3}$, and $\chi^{2}_{\rm red}$ = 1.21 ($\nu = 160$). Right: \mbox{2-T} (two-temperature) function (green and yellow curves with the red curve representing the sum of the two) fitted over the $1.5-8.5$~keV energy range with $T_1 = 22$~MK (1.9~keV), $T_2 = 2.9$~MK (0.25~keV), EM$_1= 7.6 \times 10^{48}$~cm$^{-3}$, EM$_2 = 4.3 \times 10^{49}$~cm$^{-3}$, and $\chi^{2}_{\rm red} = 1.18$ ($\nu = 175$). The same pre-flare background spectrum is shown in pink in both plots.}
    \label{fig:sp_vth}
\end{figure}

We describe the fitting procedure for the spectrum taken during the 5-minute interval beginning at 14:56:29.9~UT, at flare maximum. The observed spectrum, shown in the two panels of Figure~\ref{fig:sp_vth}, is dominated by continuum (the sum of free--free and free--bound emission), with line complexes due to Ca ions ($\sim 3.9$~keV) and Fe ions ($\sim 6.7$~keV and $\sim 8$~keV) also evident.  The temperature governs the slope of the spectrum when plotted against energy, with higher temperatures giving rise to spectra with flatter slopes, while the emission measure determines the absolute value of the flux at any energy.

The measured count rate spectrum was fitted with theoretical photon spectra calculated from the {\sc chianti} package (v.~7.0) \citep{lan13}. These spectra are defined by values of the electron temperature, $T$, and the volume emission measure, EM $= N_e^2 V$ (where $N_e = $ electron density, $V = $ emitting volume), for a set of abundances which in our analysis were chosen to be relative to the coronal abundances of \cite{fel92} with ion fractions from \cite{maz98}.
In order to speed up the calculation of the theoretical thermal spectra, base isothermal spectra were obtained from an abstraction of the {\sc chianti} routines.  Tables were generated for each {\sc chianti} spectral line for a logarithmically spaced temperature grid with the grid spacing dependent on the significance of that line.  When a particular spectrum was generated from that database, the lines were found in the energy range of interest and the emissivity computed by interpolating from the table.  The total line spectrum was then computed by summing over all line contributions with each line weighted by its assumed element abundance.  The continuum was similarly computed and made available either as a separate spectrum or combined with the total line spectrum.  In this way, EM, $T$, and the abundances of Fe, Ca, S, Si, Ar, and other elements could all be used as free parameters and adjusted as necessary in the fitting procedure.

As the temperature structure of the flare emission is unknown, functional forms must be assumed. The simplest case is an isothermal function defined only by $T$ and EM. This appears to apply most nearly to X-ray spectra over limited energy ranges taken near maximum and in the decay of a flare, as has been shown in analysis of {\em GOES} flares by \cite{gar94} and with temperature-dependent spectral line ratios from the {\em Yohkoh} Bragg Crystal Spectrometer by \cite{phi05}. In the analysis of \cite{nit11}, SAX spectra were fitted assuming an isothermal emitting region for the $1-8$~keV spectral range on the grounds that two-temperature models gave worse fits. However, isothermal fits to the entire spectral range of SAX are not likely to be valid in principle for most spectra.  This is because previous flare observations with high-resolution crystal spectrometers have shown that temperatures determined from the Ca lines at 3.9~keV (due to \ion{Ca}{19} lines with relatively weaker satellites) are found to be lower by typically $\sim 5$~MK than those determined from the Fe-line complex at 6.7~keV (due to \ion{Fe}{25} lines with temperature-dependent dielectronic satellites) \citep[e.g.][]{fel80a}. An isothermal assumption is more likely to be valid over limited spectral regions, particularly on either side of the 6.7~keV Fe-line complex, where the neighboring continuum has a slope determined by a temperature which is likely to approximate that derived from satellite-to-resonance line ratios in the \ion{Fe}{25} lines and nearby satellite lines.

For the SAX spectrum in Figure~\ref{fig:sp_vth}, four forms of the temperature dependence of the differential emission measure, defined as DEM $=N_e^2 dV/dT$, were assumed:
($a$) an isothermal model (``\mbox{1-T}'') described by single values of $T$ and EM; ($b$) a two-temperature model (``\mbox{2-T}"), with two values of $T$ and EM; ($c$) a multi-thermal model with power-law dependence of the differential emission measure on $T$ \mbox{(``multitherm\_pow"),} constant$\times T^{-\alpha}$ (where $\alpha$ is positive); ($d$) a multi-thermal model with the differential emission measure having an exponentially decreasing dependence on $T$ \mbox{(``multitherm\_exp"),} constant$\times {\rm exp} (-T / \tau)$ (where $\tau$ is positive). In our fitting procedure, theoretical spectra calculated from {\sc chianti} were folded through each of these functional forms with free parameters (e.g. $T$, EM, $\alpha$, and $\tau$) and the abundances of elements significantly affecting the spectrum in this energy range, viz., Fe, Ca, S, Si, and Ar. For the \mbox{2-T} model the abundances of the higher temperature component were allowed to vary but the abundances of the lower temperature component were held fixed at the {\sc chianti} ``coronal'' values \citep{fel92} listed in Table~\ref{tab:JuneflareFIP&FIPbias}.

The fitting routines are much the same as those used by \cite{phi12} to obtain the Fe abundance from {\em RHESSI} spectra. Data files containing the count-rate spectrum and detector response matrix were read by a program called Object Spectral Analysis Executive (OSPEX) (http://hesperia.gsfc.nasa.gov/rhessi3/software/spectroscopy/spectral-analysis-software/). Theoretical spectra calculated from each of the four model spectral functions ($a$--$d$) were then folded through the detector response function with first-guess values of the free parameters (e.g. for model ($a$) $T$, EM, and abundances of Fe, Ca, S, Si, and Ar) to calculate predicted count rates. Fits to the background-subtracted SAX count-rate spectrum (with the background determined before or after the flare) were accomplished with OSPEX by continual adjustment of the fitting parameters to improve the goodness of fit, determined by the reduced chi-squared statistic -

% Eq. 1
\begin{equation}
    \chi^{2}_{\rm red} = \frac{1}{\nu}~\sum_{i}[(C_{i, obs} - C_{i, exp})^2/\sigma_{i, exp}^2]
    \label{eq:chi2}
\end{equation}

\noindent where $\nu$ is the number of degrees of freedom (number of data points minus the number of free parameters), $C_{i,obs}$ and $C_{i,exp}$ are the measured and predicted count rates respectively in the $i$th energy bin, and $\sigma_{i,exp}$ is the standard deviation of the expected count rate assuming Poisson statistics.

% Figure 2: fits to 2007 June 1 (maximum phase) spectrum at peak with \mbox{1-T}, \mbox{2-T}, ?multi-T models

Figure~\ref{fig:sp_vth} shows the \mbox{1-T} and \mbox{2-T} fits to the measured SAX count rate flux spectrum with normalized residuals below. (Equivalent plots for the \mbox{multitherm\_pow} and \mbox{multitherm\_exp} fits are not shown but have similar appearance and all have acceptable values of $\chi^{2}_{\rm red}$.) Count rate flux is the measured count rate converted to units of counts s$^{-1}$ cm$^{-2}$ keV$^{-1}$ by dividing by the effective area projected to 1~AU and by the energy bin width.  This enables the varying distance of the {\em MESSENGER} spacecraft from the Sun to be accounted for as it passed from Earth to Mercury and during its elliptical orbit about Mercury. For the \mbox{1-T} fit, the energy range (indicated in the figure) was confined to $2.3-8.5$~keV because a good fit could not be achieved at lower energies. For the \mbox{2-T} function, a satisfactory fit was achieved over the range $1.5-8.5$~keV, i.e., including lower energies. The values of $\chi^2_{\rm red}$ for each fit are both less than 1.5, indicating a satisfactory fit to the observed spectrum in each case. For the \mbox{1-T} fit, $\chi^2_{\rm red} = 1.21$ with a probability $P~=~3.6\%$ of exceeding this value (165 energy bins each 0.0375~keV wide and five free parameters giving $\nu = 160$). For the \mbox{2-T} fit, $\chi^2_{\rm red} = 1.18$ giving $P = 5.2\%$ (187 energy bins and 12 free parameters giving $\nu = 175$).

\begin{figure}
\begin{center}
\includegraphics*[angle = 90,width = 0.8\textwidth ]{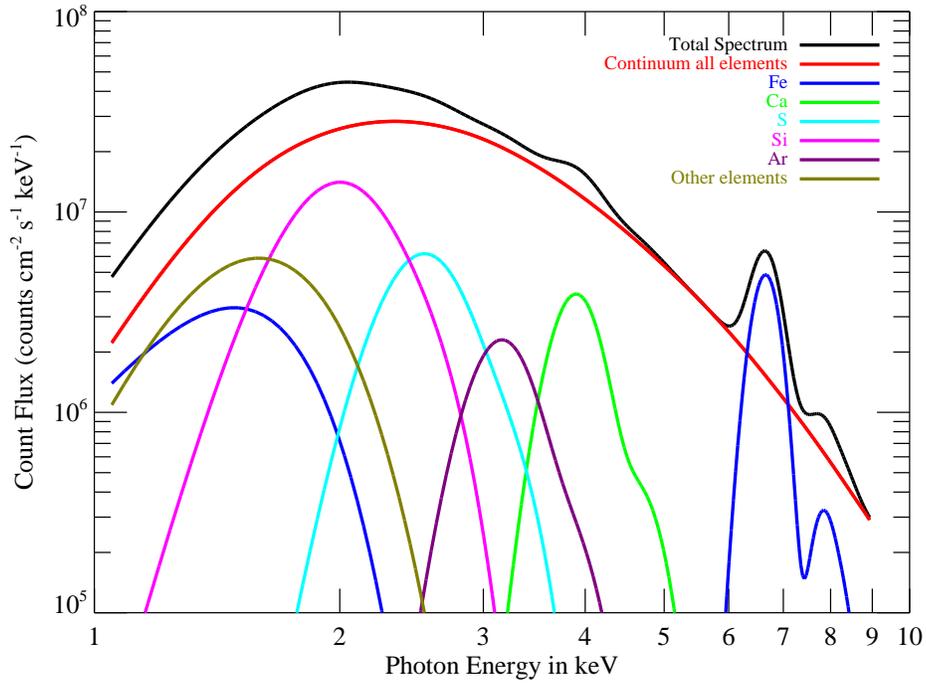}
\end{center}
\caption{Contributions to the SAX count-rate spectrum (black curve) for the interval shown in Figure~\ref{fig:sp_vth} made by the thermal continuum (red) and line complexes of elements indicated in the legend calculated using {\sc chianti} (v. 7.0) for the \mbox{1-T} model with $T = 22$~MK (1.91~keV) and with the SAX spectral resolution (598~eV at 5.9~keV).
\label{fig:sp_cont_allelems_1to10kev}}
\end{figure}

The abundances of the elements were set as free parameters and determined for each of the functional forms given above. The Fe and Ca abundance estimates are expected to have higher precision (of order $\pm20\%$) than for S, Si, and Ar since Fe and Ca spectral lines form recognizable line features in the measured spectrum (Figure~\ref{fig:sp_vth}). Separate line features from S, Si, and Ar are not evident with the SAX energy resolution but their contributions to the spectrum are considerable, as can be seen from Figure~\ref{fig:sp_cont_allelems_1to10kev}. Here, the contributions to the SAX count flux spectrum of the thermal continuum emission and the line fluxes from the different elements are shown with the SAX energy resolution and sensitivity. The photon spectrum used to calculate these different components using {\sc chianti} was the \mbox{1-T} model with a temperature of 22~MK (1.9~keV) that gave the best fit to the measured count flux spectrum in the 5-minute interval used for Figure~\ref{fig:sp_vth}. Note that the contributions from the different elements peak at the following energies: Fe -- 1.5, 6.7, and 8 keV; Ca -- 3.9 keV; S -- 2.5 keV; Si -- 2.0 keV, and Ar -- 3.2 keV. Other elements (notably Mg) give rise to emission peaking at $\sim 1.6$~keV. The SAX sensitivity is poorly known at such low energies because of the uncertain absorption of overlying material in the instrument and the presence of a steep and time-varying background spectrum, so spectral fits were limited to energies above 1.5~keV and no attempt was made to determine the Mg abundance.

As mentioned above, the continuum is made up of free--free and free--bound radiation, and so varying element abundances in the spectral fits in principle affects free--bound emission. \cite{phi12} found that varying the Fe abundance  from the (e.g.) photospheric \citep{asp09} to coronal \citep{fel92} values leads to only a few percent difference in the total continuum at 10~keV for a temperature of 25~MK; for the lower temperatures and energies  mostly appropriate here, the effect will be less and can be ignored.

% Figure 3

Estimates of abundances and FIP biases of Fe, Ca, S, Si, and Ar from the two model fits to the spectrum of Figure~\ref{fig:sp_vth} are given in Table~1.
% along with the energy ranges used for the fits and the $\chi^2_{\rm red}$ values.
Also given are the corresponding FIP biases from the abundance sets of \cite{fel92} and \cite{flu99}.

% Table 1: List of Messenger element abundances
\begin{deluxetable}{cccccccc}
\tabletypesize{\scriptsize} \tablecaption{Abundances and FIP biases from the SAX spectrum of the 2007 June~1 flare (14:56:30--15:01:29~UT)}

\tablewidth{0pt}
\tablehead{\colhead{Element} & \colhead{FIP} & \multicolumn{4}{c}{Fitting Model (see text)} & \colhead{Photo-} & \colhead{Corona$^b$} \\
        & \colhead{(eV)}         & \colhead{1-T} & \colhead{\mbox{2-T}} & \colhead{$aT^{-\alpha}$} & \colhead{$a~{\rm exp} (-T / \tau)$} & \colhead{sphere$^a$}\\ }
\startdata
Fe & 7.87 & 2.33$^c$  [7.87$^d$] & 1.70 [7.73] & 1.22 [7.59] & 1.76 [7.74] & 7.50$\pm$0.04 & 3.98 [8.10] \\
Ca & 6.11 & 3.89 [6.93] & 3.59 [6.90] & 3.17 [6.84] & 2.98 [6.81] & 6.34$\pm$0.04 & 3.89 [6.93] \\
S & 10.36 & 0.96 [7.10] & 0.78 [7.01] & 0.46 [6.78] & 0.45 [6.77] & 7.12$\pm$0.03 & 1.41 [7.27] \\
Si & 8.15 & 2.04 [7.82] & 1.09 [7.55] & 0.69 [7.35] & 0.75 [7.39] & 7.51$\pm$0.03 & 3.89 [8.10] \\
Ar & 15.76 & 0.85 [6.33] & 1.48 [6.57] & 0.71 [6.25] & 0.37 [5.97] & 6.40$\pm$0.13 & 1.51 [6.58] \\
% K & 4.34 &  &  &  &  & 5.03$\pm$0.09 & 4.37 [5.67] \\
\enddata
\tablenotetext{a} {Photospheric abundances from \cite{asp09} on logarithmic scale with $A({\rm H}) = 12$.}
\tablenotetext{b} {{\sc chianti} ``coronal abundances'' from \cite{fel92}.}
\tablenotetext{c} {FIP bias with uncertainties (abundances on logarithmic scale in square brackets).}
\tablenotetext{d} {Abundances on a logarithmic scale with $A({\rm H}) = 12$.}
\label{tab:JuneflareFIP&FIPbias}
\end{deluxetable}

% Figure 4
\begin{figure}
    \begin{center}
    \includegraphics*[angle = 0, width = 0.5\textwidth, ]{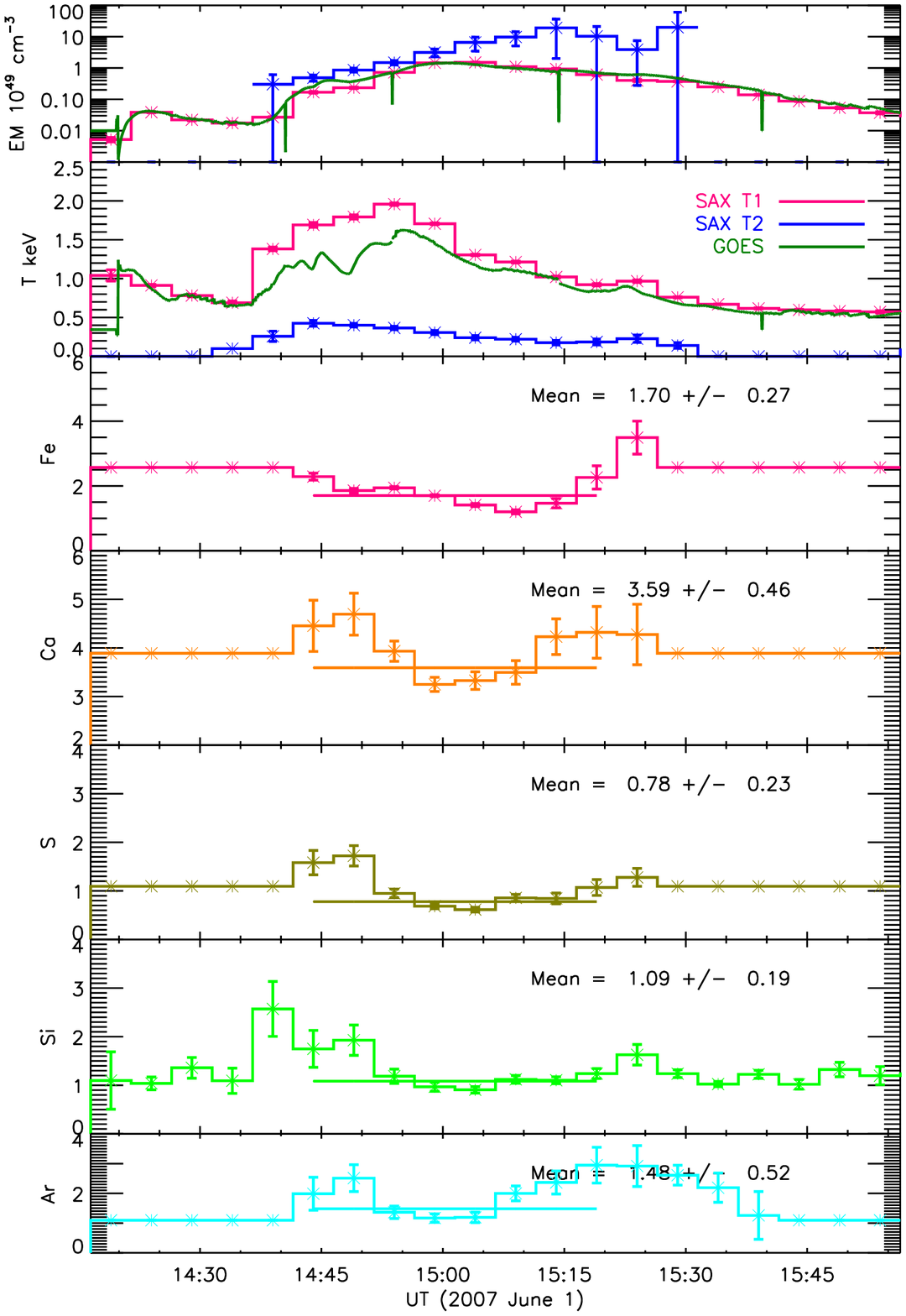}
    \end{center}
    \caption{Estimated emission measures, temperatures, and abundances (with $\pm 1\sigma$ uncertainties) from SAX spectra for the 2007 June~1 flare (maximum 14:59~UT) from the two-temperature (\mbox{2-T}) fit function. Top two panels: emission measures and temperatures (keV) determined from SAX (red and black) and from {\em GOES} (green). Panels 3--7 from top: FIP biases for Fe, Ca, S, Si, and Ar. The horizonal lines show the weighted average abundance estimates over the indicated time interval, and these estimates with uncertainties are indicated in the legend. }
    \label{fig:lc_SAX_EM_T_Fe_Ca}
\end{figure}

Figure~\ref{fig:lc_SAX_EM_T_Fe_Ca} shows the temperatures, emission measures, and estimated abundances of Fe, Ca, S, Si, and Ar for all time intervals using the \mbox{2-T} fit procedure over the duration of the 2007 June~1 flare. The temperatures and emission measures derived from the  flux ratio of the two {\em GOES} channels are also shown. From 14:30 to 15:30 UT, there is a hotter component with a temperature generally above 11.6~MK (1~keV) and a cooler component with a temperature below 6~MK (0.5~keV). Apart from the first few time intervals of the flare during its impulsive phase, the {\em GOES} temperature is only slightly less than the hotter component (labeled ``SAX T1'' in Figure~\ref{fig:lc_SAX_EM_T_Fe_Ca}).

The abundance values for all elements in Figure~\ref{fig:lc_SAX_EM_T_Fe_Ca} show variations apparently larger than the statistical uncertainties given by the OSPEX routine. These uncertainties are obtained from the error matrix and correspond to $\chi^{2}_{\rm red} + 1$ values \citep{bev03}. However, they do not take into account the interdependencies of the different parameters, which have been recently discussed by \cite{ire13}, and hence should be considered as lower limits. The horizontal line in the plot for each element shows the weighted average abundance for the time interval represented by the line length when the count flux spectrum could be measured over the full energy range.

% Subsection 2.3
\subsection{Comprehensive analysis of SAX Flares 2007 -- 2013}
\label{companal}

The detailed analysis of the 2007 June 1 flare above was generalized to spectra taken in flares detected by SAX in the 2007--2013 period and selected according to criteria given in Section~\ref{flareselection}. Care was taken to avoid selection biases and effects of varying instrument performance and the changing environment in which {\em MESSENGER} was located, both in its cruise phase and in orbit about Mercury. A semi-automatic routine was written in the Interactive Data Language (IDL) to analyze all the larger flares detected with SAX up to 2013 September~13 with the same basic techniques described using the \mbox{1-T} and \mbox{2-T} temperature functions. (The \mbox{multitherm\_pow} and \mbox{multitherm\_exp} functions were not used at this stage since acceptable fits were generally achieved with the \mbox{1-T} or \mbox{2-T} models.) The procedure handled the variations in the performance and background spectrum over this extensive period, analyzing spectra in each time period when the count rate was above a given threshold.

The data were analyzed in three stages. First, a variable background spectrum was determined for each 24~hr period of the observations.  This was achieved by selecting the 10\% of all the observational time periods with the lowest count rates in the $1.3-5.3$~keV energy range and using those values to determine a smoothed time variation by applying a \cite{sav64} filter with a width of 128 intervals.  This same time profile, normalized by the average rate in the background intervals, was used to estimate the background flux during a flare interval for each energy bin. (See http://hesperia.gsfc.nasa.gov/ssw/packages/spex/doc/ospex\_explanation.htm\#Background for details.)  The background spectrum estimated in this way was then subtracted from the measured spectrum in each flare time interval for analysis in OSPEX.  Only those background-subtracted spectra having count rates in the $6.3-7.0$~keV range of $>5000~s^{-1}$ were selected for analysis, this being the limit at which the Fe-line complex at 6.7~keV could be reliably measured.

In the second stage, OSPEX was used to analyze the background-subtracted count flux spectrum for each selected time interval using both the \mbox{1-T} and \mbox{2-T} temperature models.  A FITS file was generated for each model covering an entire one-year period of observations, containing the count flux spectra and best-fit model spectra with all parameters for each analyzed time interval.

For the \mbox{2-T} model, the spectral fit was performed over the energy range $1.5-8.5$~keV as with the analysis of the 2007 June~1 flare discussed above. This range avoids both low energies ($< 1.5$~keV) with a variable instrumental component (see Figure~\ref{fig:sp_vth}) and high energies with variable background count rates from charged particles. The free parameters include two emission measures (EM$_1$ and EM$_2$) and temperatures ($T_1$ and $T_2$ where $T_1~<~T_2$) as well as abundances of Fe, Ca, S, Si, and Ar (in order of decreasing measurement certainty). For 35\% of the time intervals, the emission measure of the cooler ($T_1$) component fell to values of $<10^{43}$~cm$^{-3}$, whereupon the fitting routine automatically set its value to zero and continued fitting with a single temperature and emission measure. A pseudo-function (drm\_mod) was also included to permit the automatic adjustment of the SAX nominal energy resolution and energy calibration. The instrument energy calibration can be found from the recognizable peaks at the known energies of the Fe and Ca line complexes, allowing both the intercept and slope of the energy calibration function to be estimated. The observed width of the Fe-line complex gives the detector resolution (FWHM, as a fraction of the measured pre-launch value). The number of free parameters in this case is therefore 12 (10 when $EM_1$ was set to zero). Some 186 energy bins (each 0.0375~keV wide) are included in the $1.5-8.5$~keV fit energy range, giving $\nu = 174$.

For the \mbox{1-T} case, we performed spectral fits over the energy range  $5.5-8.5$~keV (80 energy bins), which is similar to the range used for the fits to {\em RHESSI} spectra done in the analysis by \cite{phi12}. In this range, only the Fe-line complexes at 6.7~keV and 8~keV are included, so only the Fe abundance can be estimated (ionized Ni lines make a small contribution to the 8~keV complex). Also, since the Ca line complex is not included, only the resolution and intercept of the instrument's energy calibration function can be determined. These, with $T$, EM, and the Fe abundance, are the free parameters, so  $\nu = 75$.

Once the FITS files were made for all spectra and flares, selection and statistical analysis of the spectra were made in a third and final stage. Several criteria were used for including those spectra that were free from unwanted environmental, instrumental, or other effects. Thus, spectra were only included if: ($a$) they were in flare decay periods (those in the initial stages of flares had evidence of time-varying element abundances as suggested by the plots in Figure~\ref{fig:lc_SAX_EM_T_Fe_Ca}); ($b$) for the \mbox{1-T} case, temperatures were in the range $12.8 - 23.2$~MK ($1.1-2.0$~keV) (at temperatures $\lesssim 12.8$~MK the emission function of the Fe 6.7~keV line complex falls sharply so the measured intensity of the line above the continuum is uncertain, while spectra having measured temperatures $\gtrsim 23$~MK were nearly always the result of charged particle contamination of the X-ray spectrum); ($c$) emission measure EM $ > 10^{48}$~cm$^{-3}$ (lower emission measures gave rise to count rates too low to obtain reliable abundance estimates); ($d$) there were at least two contiguous 5-minute time intervals; ($e$) the values of $\chi^2_{\rm red}$ of the fitted spectra were in the range $0.5-1.5$ (values outside this range indicated poor fits or low count rates); ($f$) there was no evidence of charged particle contamination (identified by the count rate ratio in the $9-9.5$~keV to $8-8.5$~keV ranges being larger than 20).

%%%%%%%%%%%%%%%%%%%%%%%%%%%%%%%%%%%%%%%%%%%%%%%%%%%%%%%%%%%%%%%%%%%%%%%%%%%%%%%%%%%%%%%%%%%%%%%%%%%%%%%%%%%%%%%%%%%%%%%
% Section 3
\section{RESULTS OF COMPREHENSIVE ANALYSIS}
\label{results}

We summarize the results of the comprehensive analysis of 526 flares observed with SAX between 2007 May and 2013 August. The Fe abundances are presented first as determined from the \mbox{1-T} spectral fits over a narrow energy range, and these are followed by the abundances of Fe, Ca, Si, S, and Ar determined from the \mbox{2-T} model fits over a broader energy range.  (As indicated in Section~\ref{flareselection}, the results and plots are given in
http://hesperia.gsfc.nasa.gov/messenger/results/plots\_and\_tables/.)

% Subsection 3.1
\subsection{Fe abundance estimates using the \mbox{1-T} function}
\label{1-T_section}

\begin{figure}
    \begin{center}
        \includegraphics*[angle = 90, width = 0.8\textwidth, trim = 60 0 20 0] {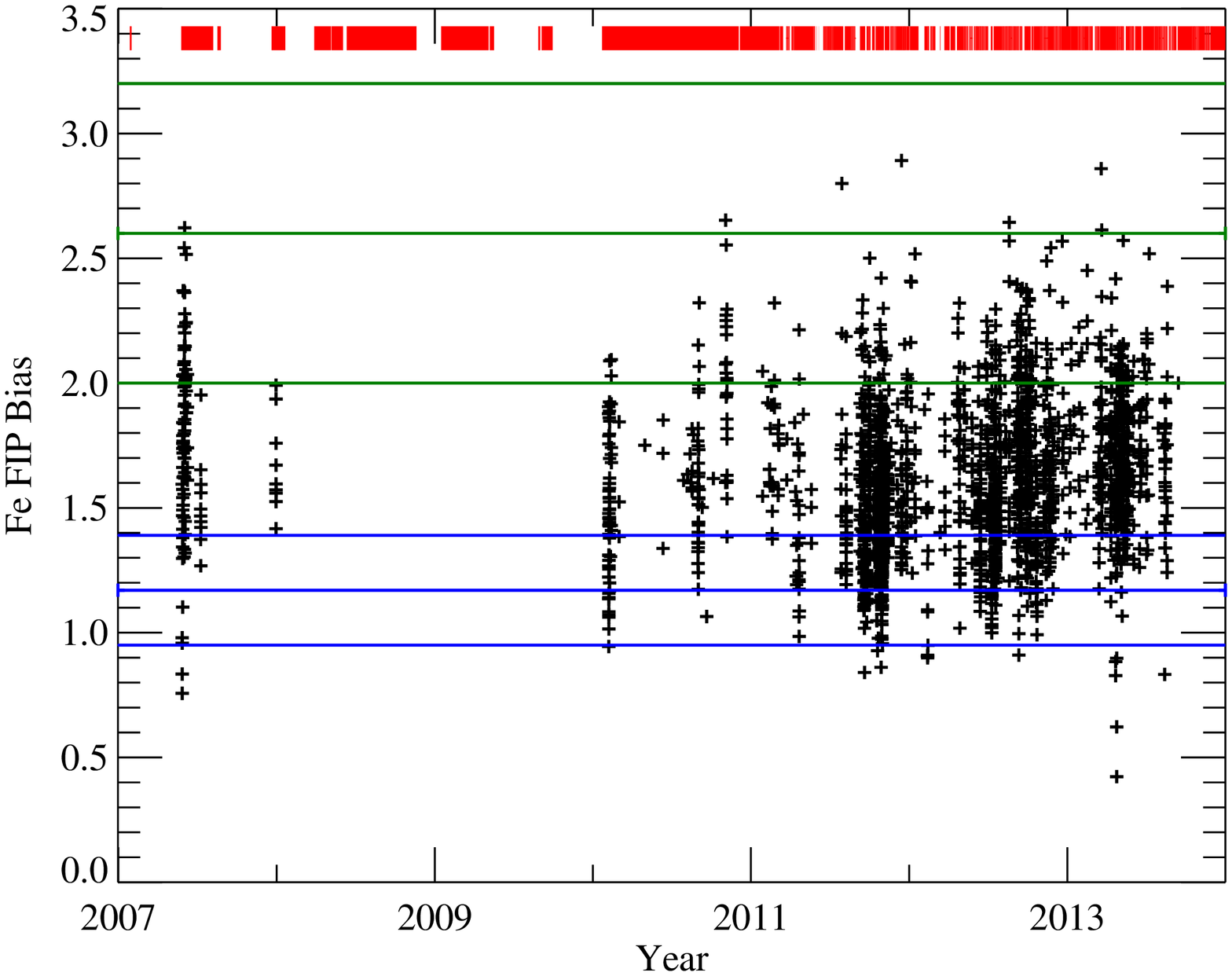}
        \includegraphics*[angle = 90, width = 0.8\textwidth, trim =  0 0 20 0] {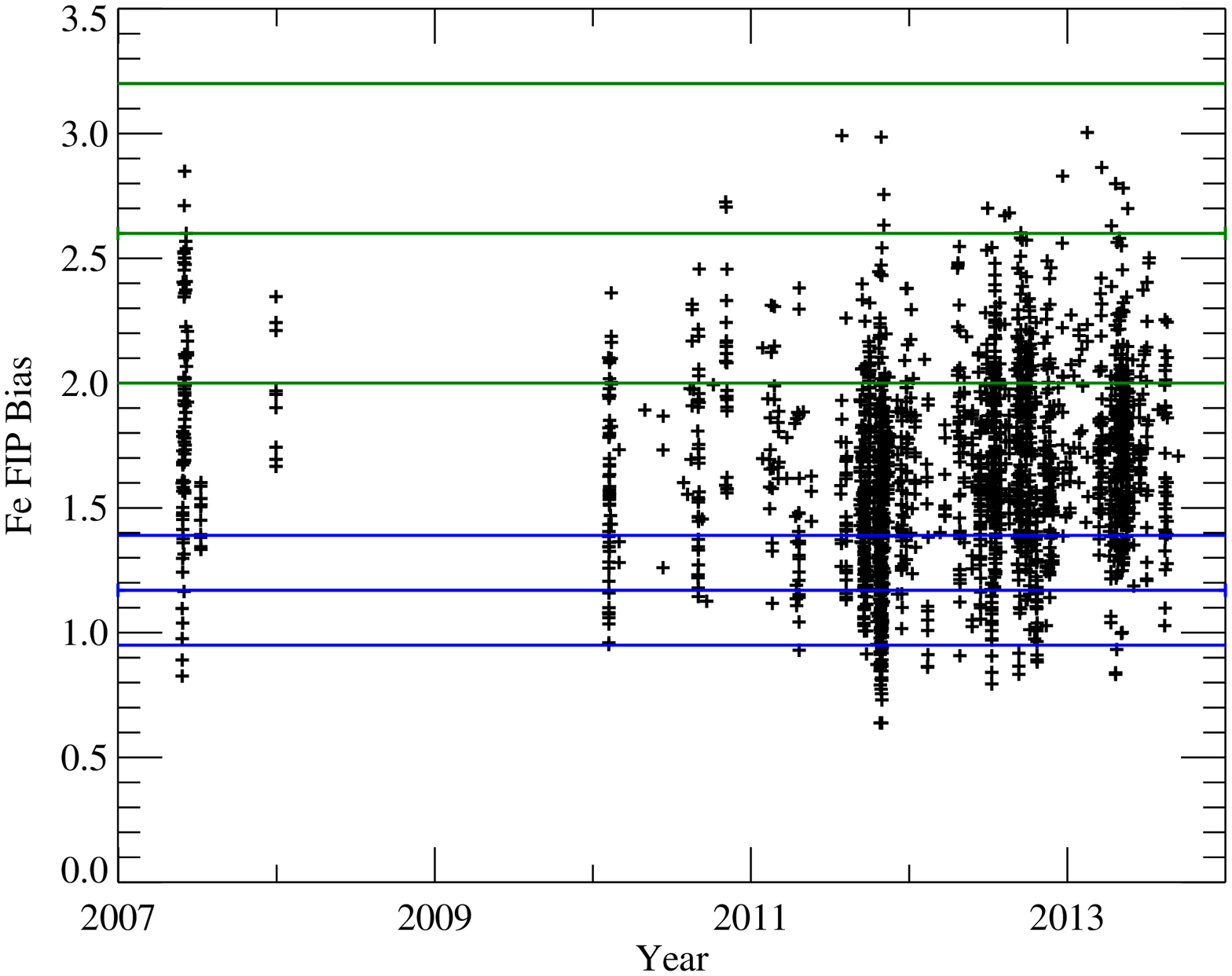}
    \end{center}
    \caption{Top: Fe FIP bias plotted against time estimated from spectral fits with the \mbox{1-T} function in the energy range $5.5 - 8.5$~keV. Bottom: FIP bias over the same period from spectral fits with the \mbox{2-T} function in the energy range $1.5-8.5$~keV. The broken red line at the top  shows the SAX data coverage. The green and blue horizontal lines show the mean FIP bias with $\pm1\sigma$ uncertainties given by \cite{phi12} and \cite{war14}, respectively.}
    \label{fig:Fe_vs_time}
\end{figure}

\begin{figure}
    \begin{center}
        \includegraphics*[angle = 90, width = 0.5\textwidth, trim = 60 0 20 0] {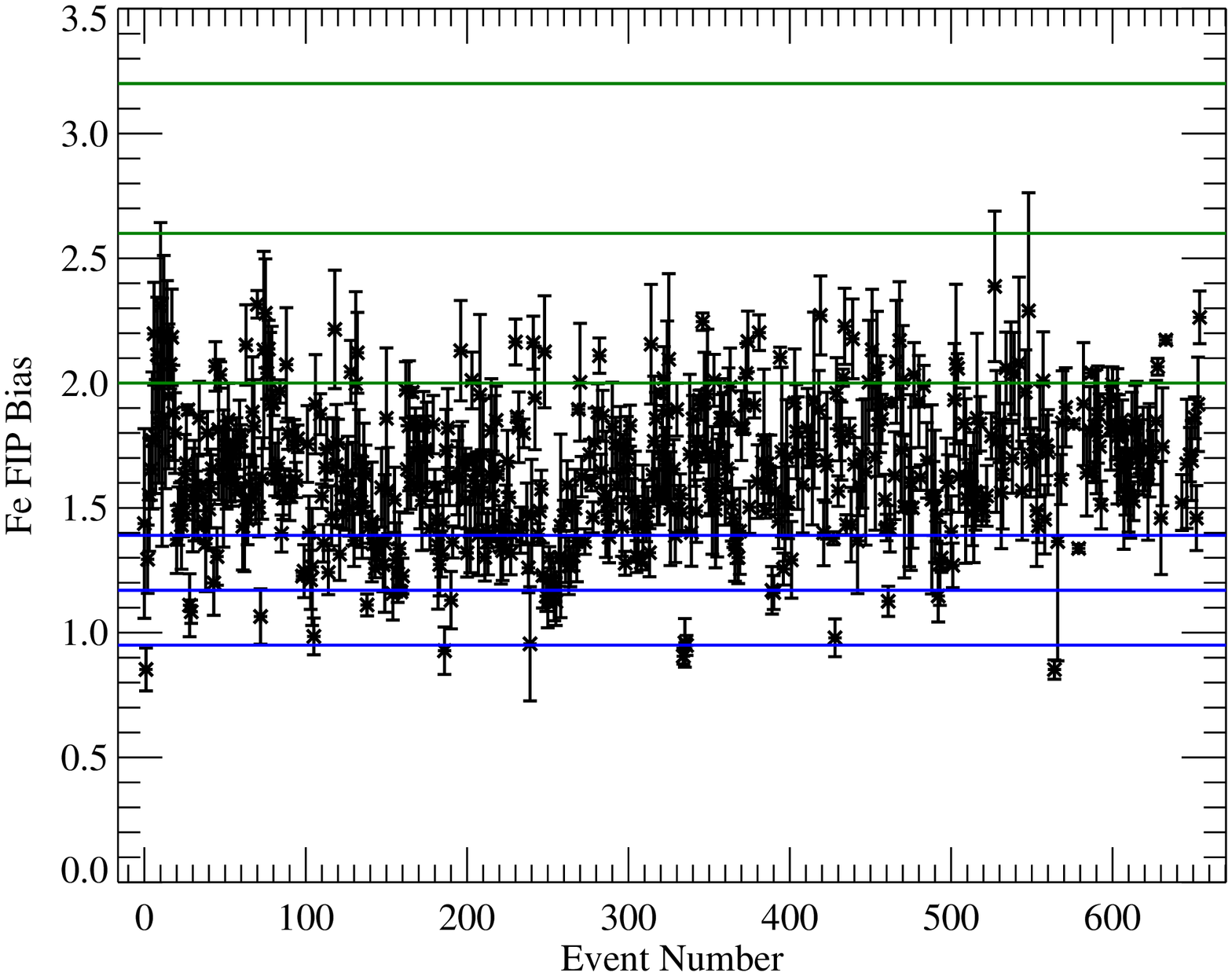}
        \includegraphics*[angle = 90, width = 0.5\textwidth, trim = 60 0 20 0] {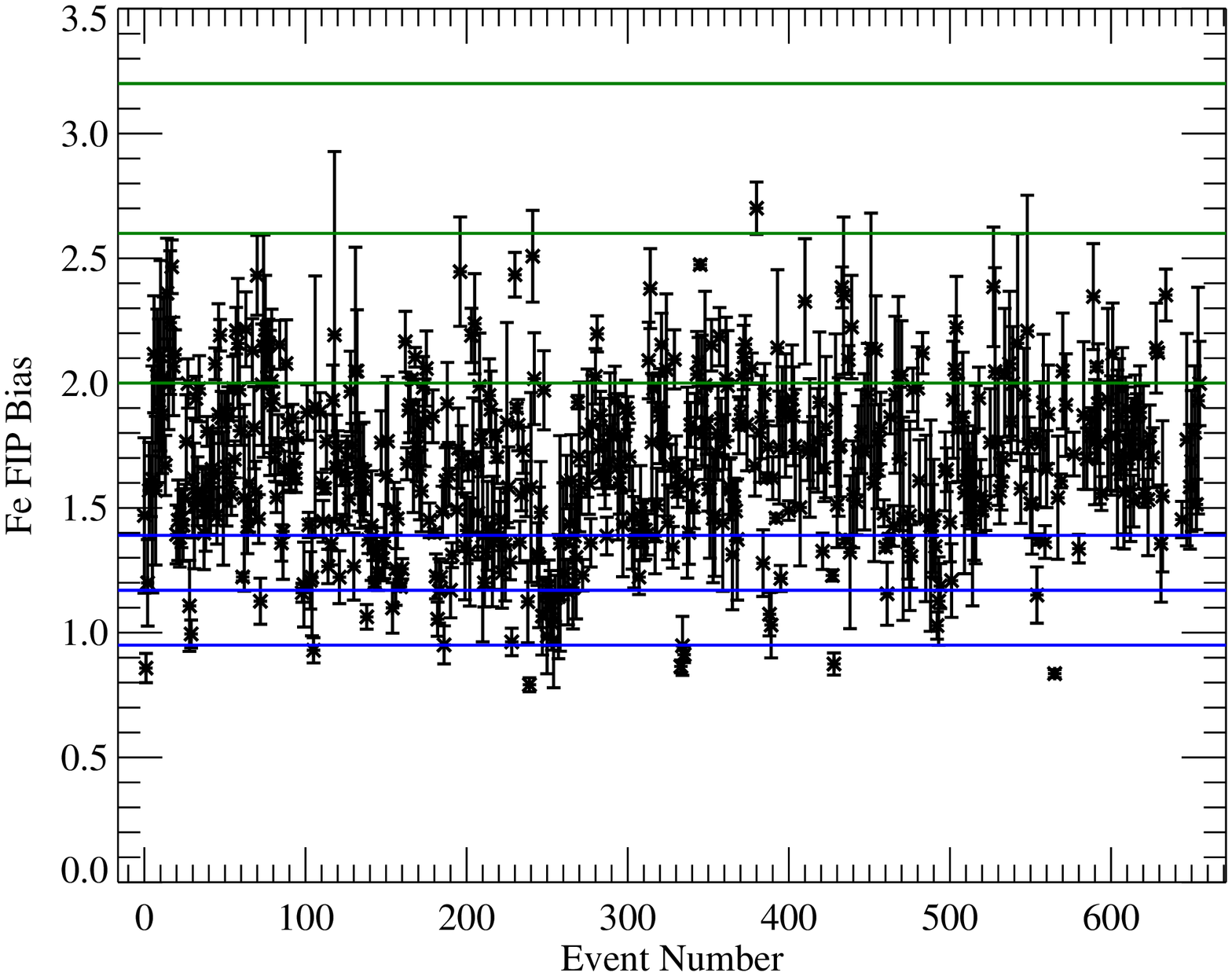}
        \includegraphics*[angle = 90, width = 0.5\textwidth, trim =  0 0 20 0] {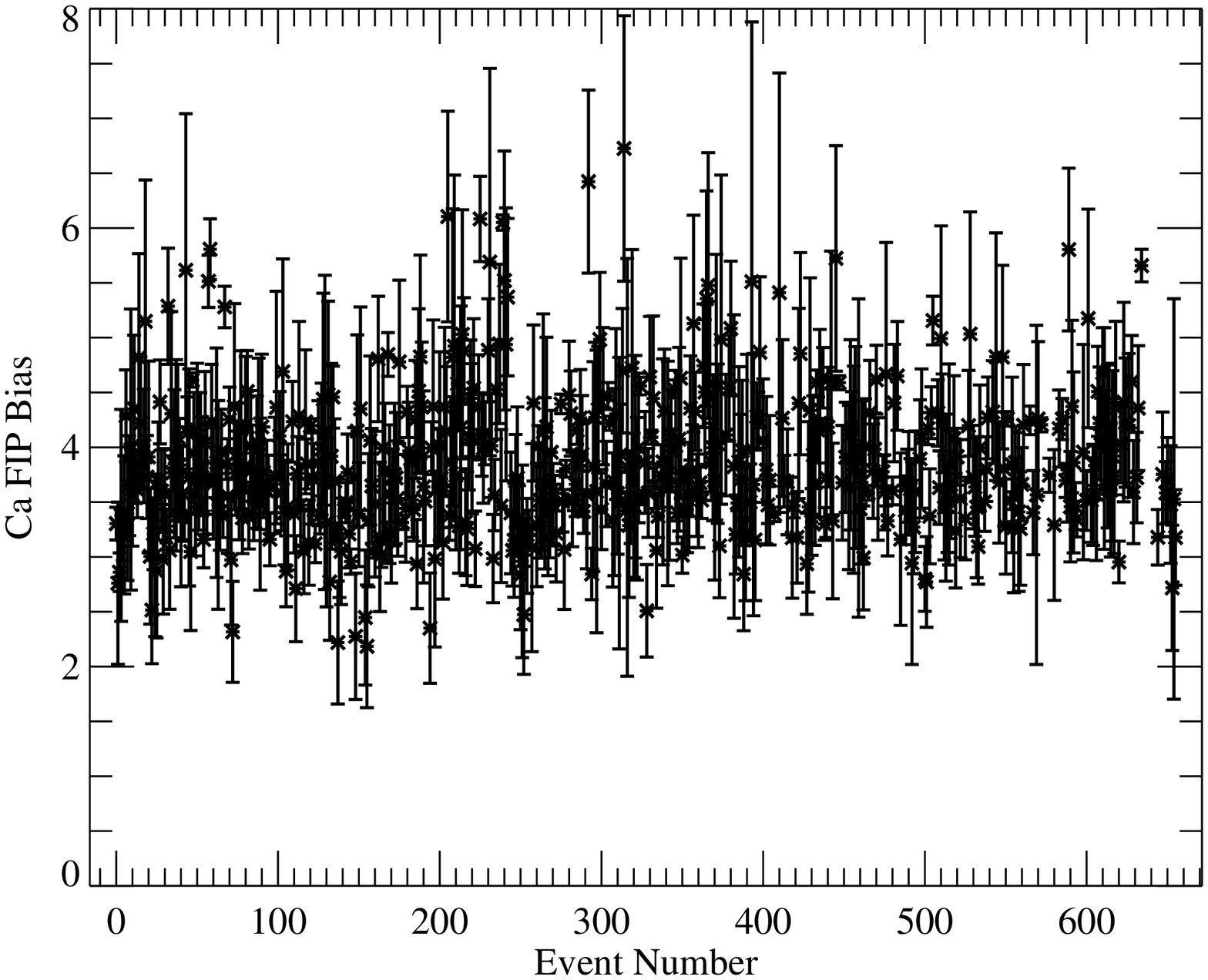}
    \end{center}
    \caption{Top: Fe FIP bias determined from \mbox{1-T} function fits (fit range $5.5-8.5$~keV) averaged over each flare plotted against flare sequence number. The green and blue horizontal lines show the mean FIP bias and $\pm1\sigma$ uncertainties given by \cite{phi12} and \cite{war14} respectively. Middle: same as the top plot with  \mbox{2-T} function fits (fit range $1.5-8.5$~keV) with the abundances of other elements set free.  Bottom: same as the middle plot but for the Ca FIP bias.}
    \label{fig:Fe1vsENo}
\end{figure}

The Fe FIP bias estimates for all spectra analyzed with the \mbox{1-T} function over the energy range $5.5-8.5$~keV are plotted against time in Figure~\ref{fig:Fe_vs_time} (top panel). The period analyzed covered the declining portion of Cycle~23 (2007--2008) and the rising portion of Cycle~24 (2010--2013). With fewer free parameters, the \mbox{1-T} fits are likely to be more reliable than those from the \mbox{2-T} function (Section~\ref{2-T_section}). Plotted in Figure~\ref{fig:Fe1vsENo} (top panel) are FIP bias values plotted against flare sequence number, with the FIP bias averaged over the duration of each flare. This plot shows the existence of flare-to-flare variations since the scatter of points is significantly larger than the uncertainties on individual points.  The uncertainties are a combination of the uncertainties given by the OSPEX routine and variations between different times during the same flare. There is at least one group of flares, flare numbers 238--261 occurring between 2011 October~28 and 30, that have mostly smaller-than-average Fe abundances. These flares may all come from a single active region although {\em MESSENGER} was then viewing the solar hemisphere partly directed away from the Earth so we cannot be sure of this. {\em STEREO B} was viewing the Sun from a similar angle to {\em MESSENGER} at that time and showed an active region at a Stonyhurst heliographic longitude of $\sim-100^{\circ}$ that could have been the origin of at least some of these flares. None of them are evident in the {\em GOES} light curves except for six on 2011 October 28 all at the C1 to C2 level.

% Figure 5 - Fe FIP bias vs. time 2007 - 2013

% Figure 6 - FIP bias vs. flare number

\begin{figure}
    \begin{center}
    \includegraphics*[angle = 90, width = 0.5\textwidth, trim = 60 0 20 0] {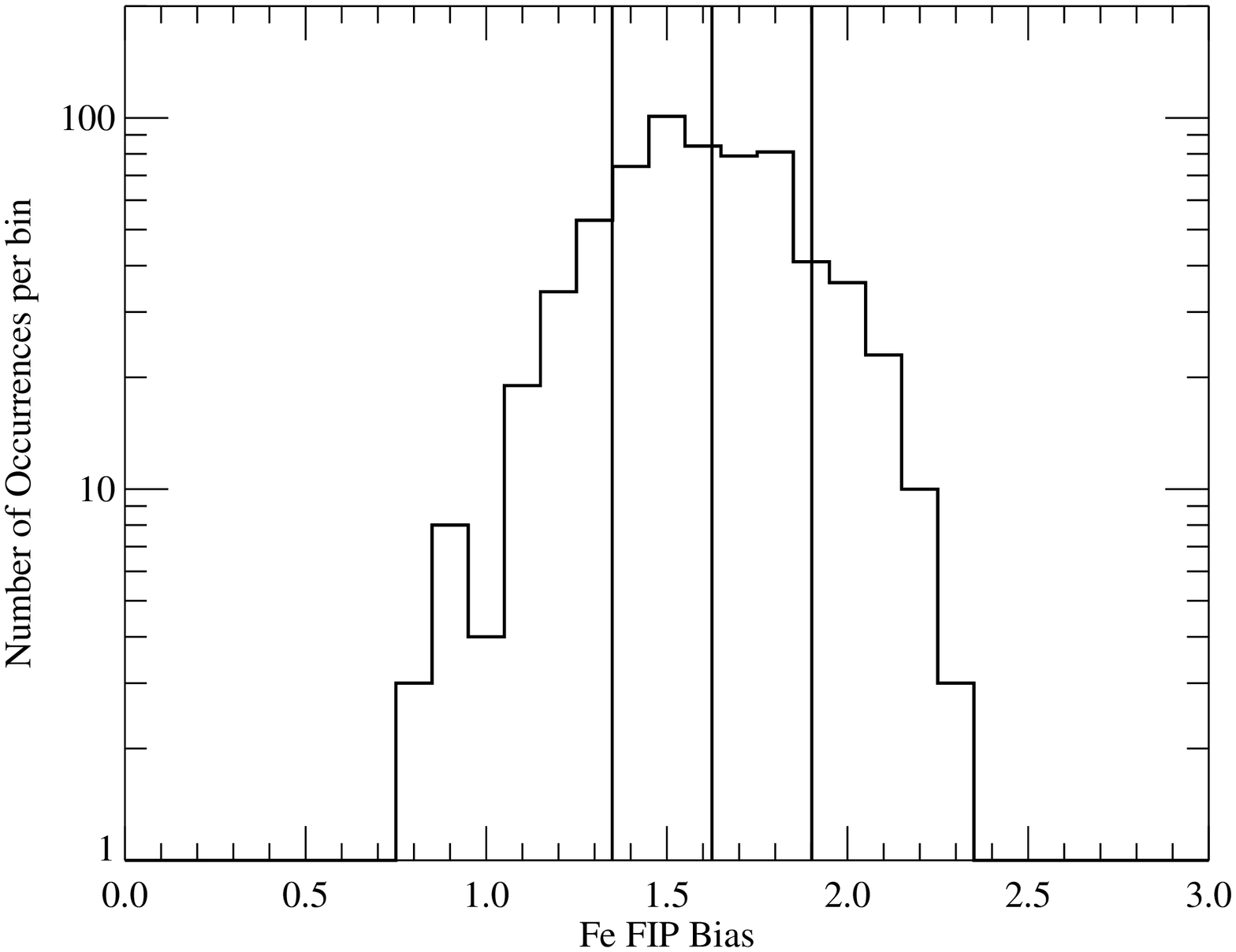}
    \includegraphics*[angle = 90, width = 0.5\textwidth, trim = 0 0 20 0 ] {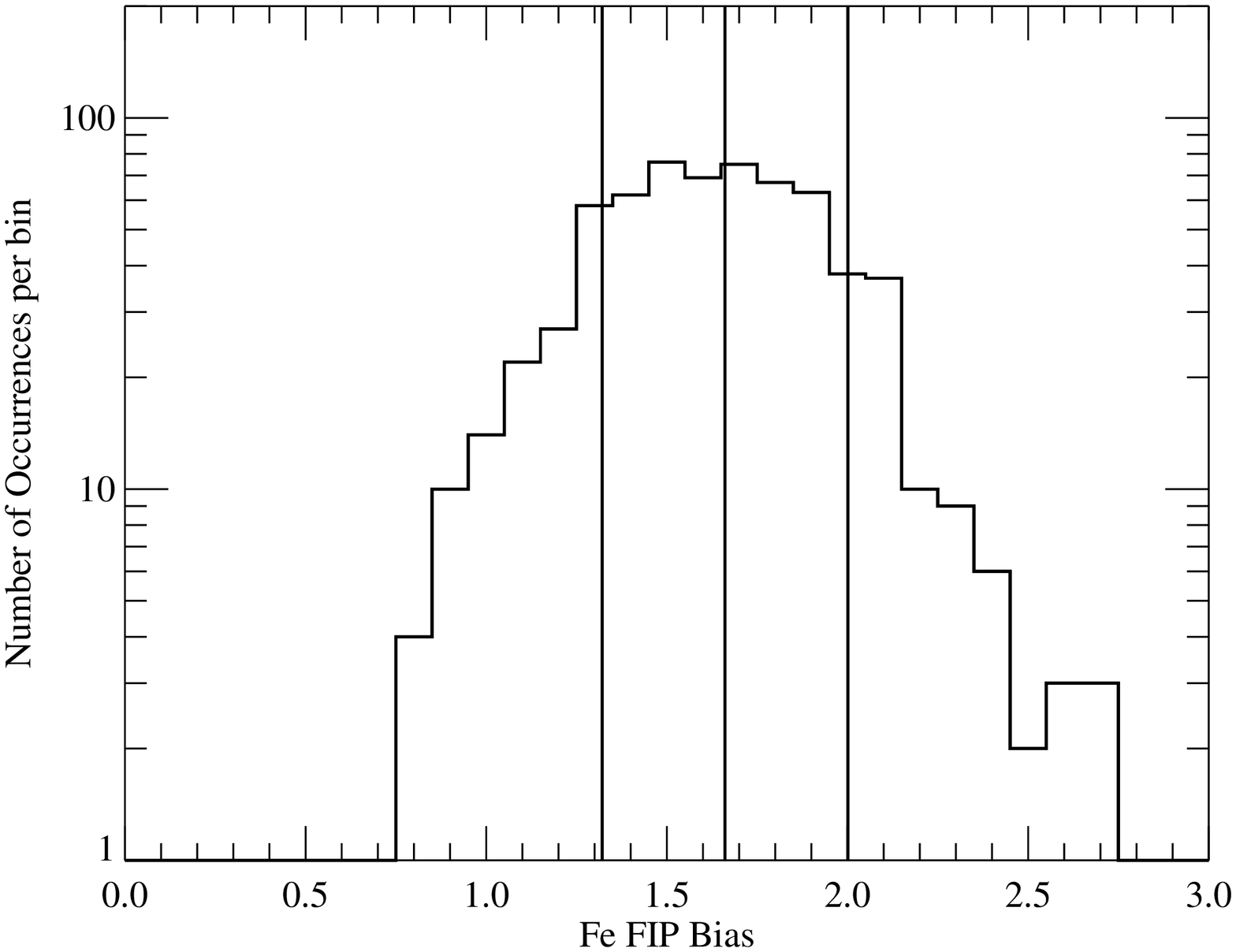}
    \end{center}
    \caption{Frequency histograms for the Fe FIP bias values determined from fits using the \mbox{1-T} function (top: energy range $5.5-8.5$~keV) and \mbox{2-T} function (bottom: $1.5-8.5$~keV) for all selected time intervals. The mean FIP bias is $1.62 \pm 0.28$ (s.d.) for the \mbox{1-T} distribution and $1.66 \pm 0.50$ for the \mbox{2-T} distribution.}
    \label{fig:OccvsFe1}
    \end{figure}

Figure~\ref{fig:OccvsFe1} (top panel) shows the frequency distribution of FIP bias estimates obtained for all spectra. From this distribution, the mean FIP bias from the \mbox{1-T} function is $1.62 \pm 0.28$ (s.d.), corresponding to an Fe abundance (on a logarithmic scale with H = 12) of $A({\rm Fe}) = 7.71 \pm 0.07$. The standard deviations include any real time variations over the period of observations.

% Figure 7 - Frequency histograms for Fe abundance estimates

\begin{figure}
    \begin{center}
        \includegraphics*[angle = 90, width = 0.5\textwidth, trim = 0 0 20 0] {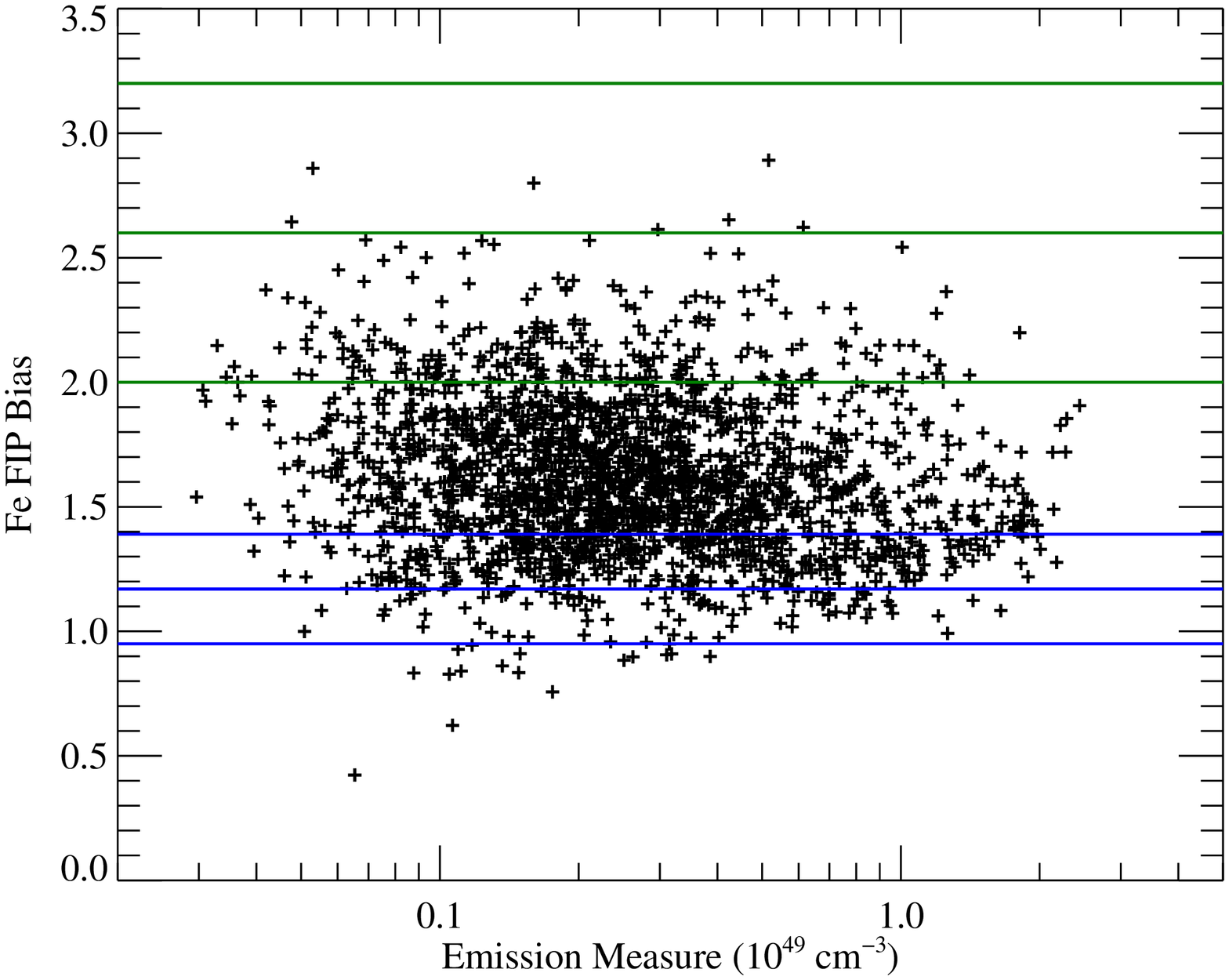}
        \includegraphics*[angle = 90, width = 0.5\textwidth, trim = 0 0 20 0] {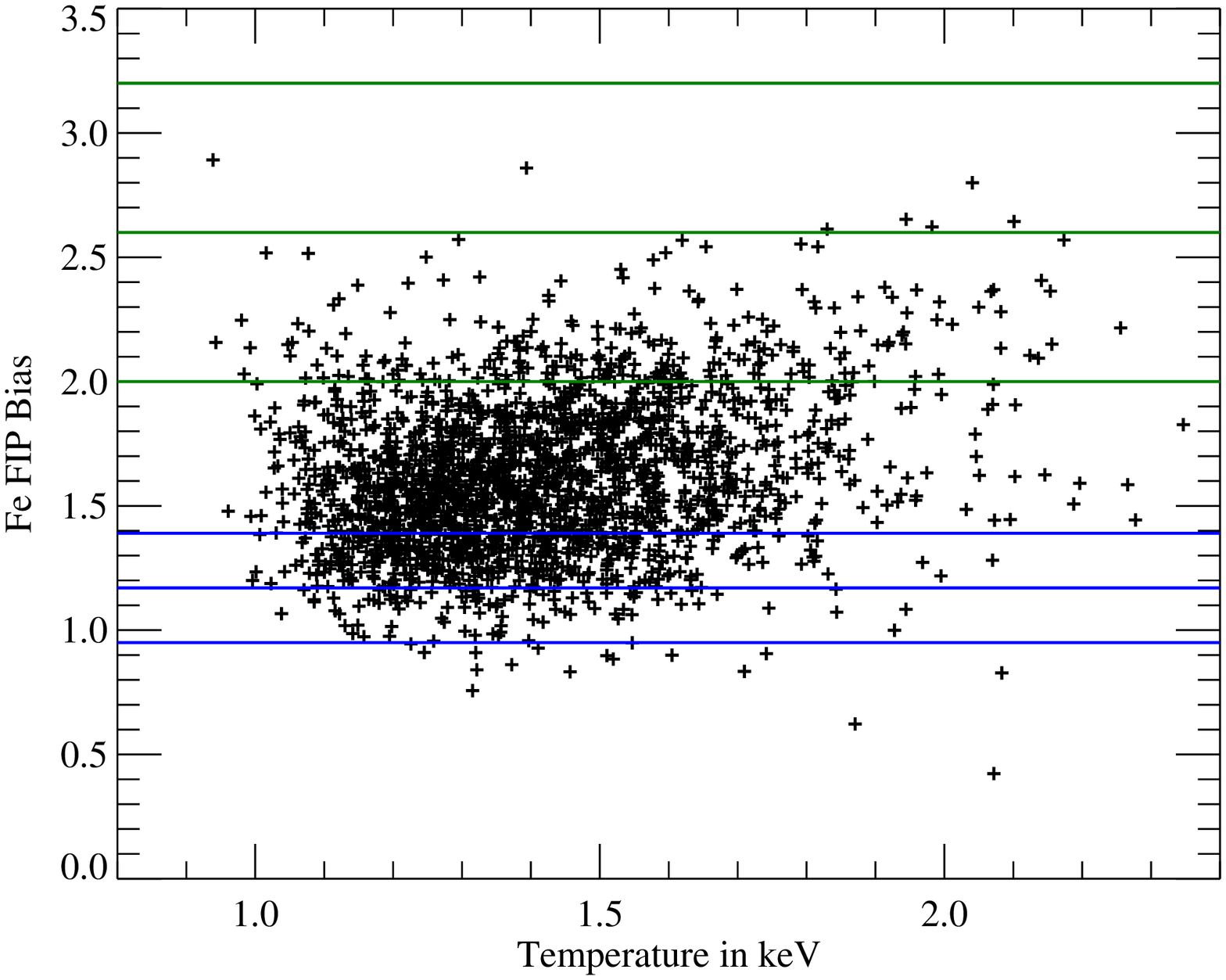}
    \end{center}
    \caption{Fe FIP bias estimates from the \mbox{1-T} fitting function (energy range $5.5-8.5$~keV) from all spectra plotted against emission measure (top) and against temperature (bottom).
%     Bottom plot: Fe FIP bias estimates from the \mbox{2-T} fitting function (energy range $1.5-8.5$~keV) against the emission measure of the hotter component. Same as the top plot but for the \mbox{2-T} function fitted between 1.5 and 8.5 keV. The FIP bias is plotted vs.~the emission measure of the higher temperature component.
     In both plots, the green and blue horizontal lines show the mean FIP bias with $\pm 1\sigma$ uncertainties given by \cite{phi12} and \cite{war14}, respectively.}
    \label{fig:Fe1vsEM}
\end{figure}

To check for any trends with flare parameters, the Fe FIP bias is plotted against both emission measure and temperature in  Figure~\ref{fig:Fe1vsEM}. There appears to be a slight tendency for lower FIP biases to result from larger emission measures ($\gtrsim 10^{49}$~cm$^{-3}$) and lower temperatures, but the significance of the trend is low.

% Figure 8 - Fe FIP bias vs. EM and T2

\begin{figure}
    \begin{center}
       \includegraphics*[angle = 90, width = 0.5\textwidth, trim = 60 0 20 0 ] {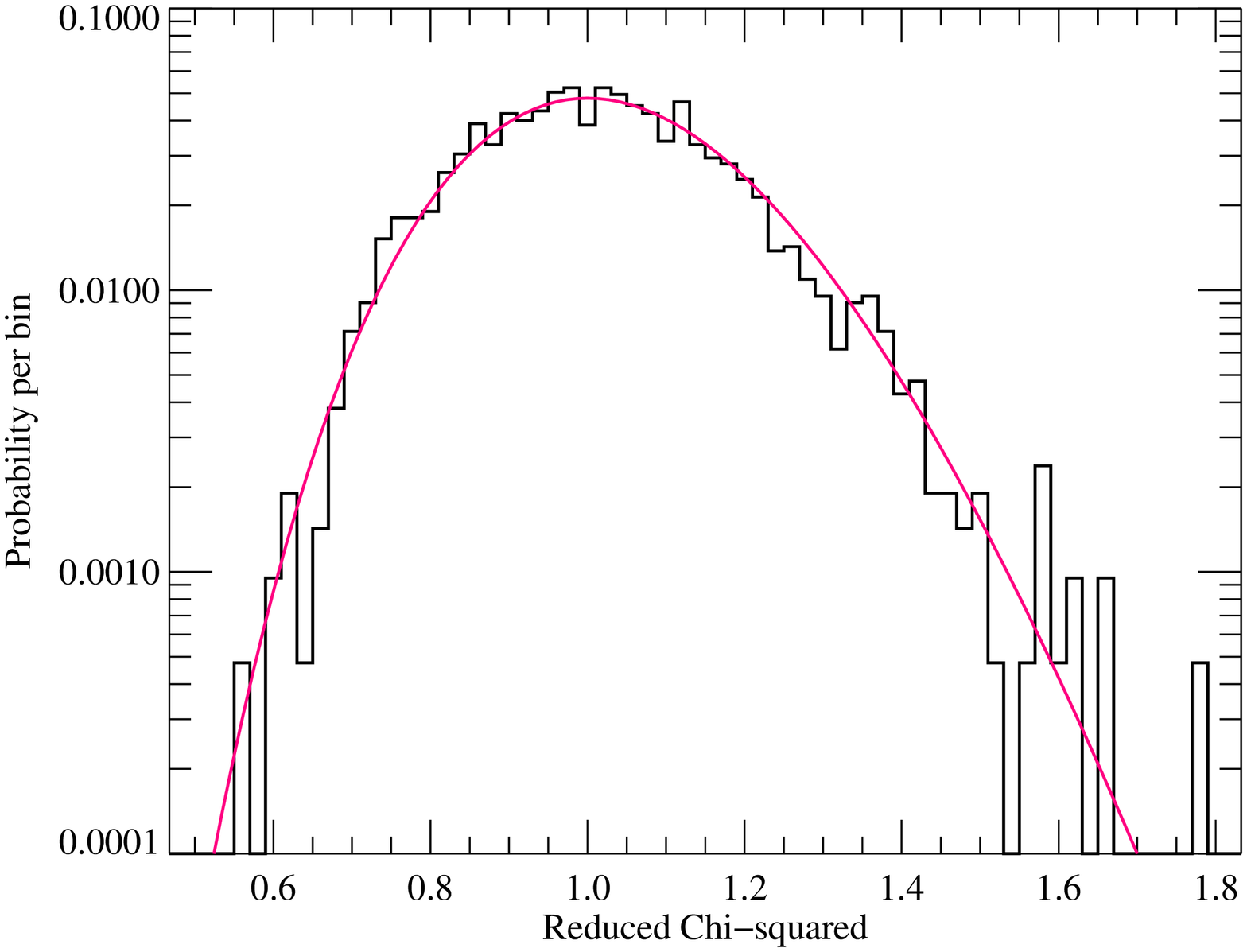}
       \includegraphics*[angle = 90, width = 0.5\textwidth, trim =  0 0 20 0 ] {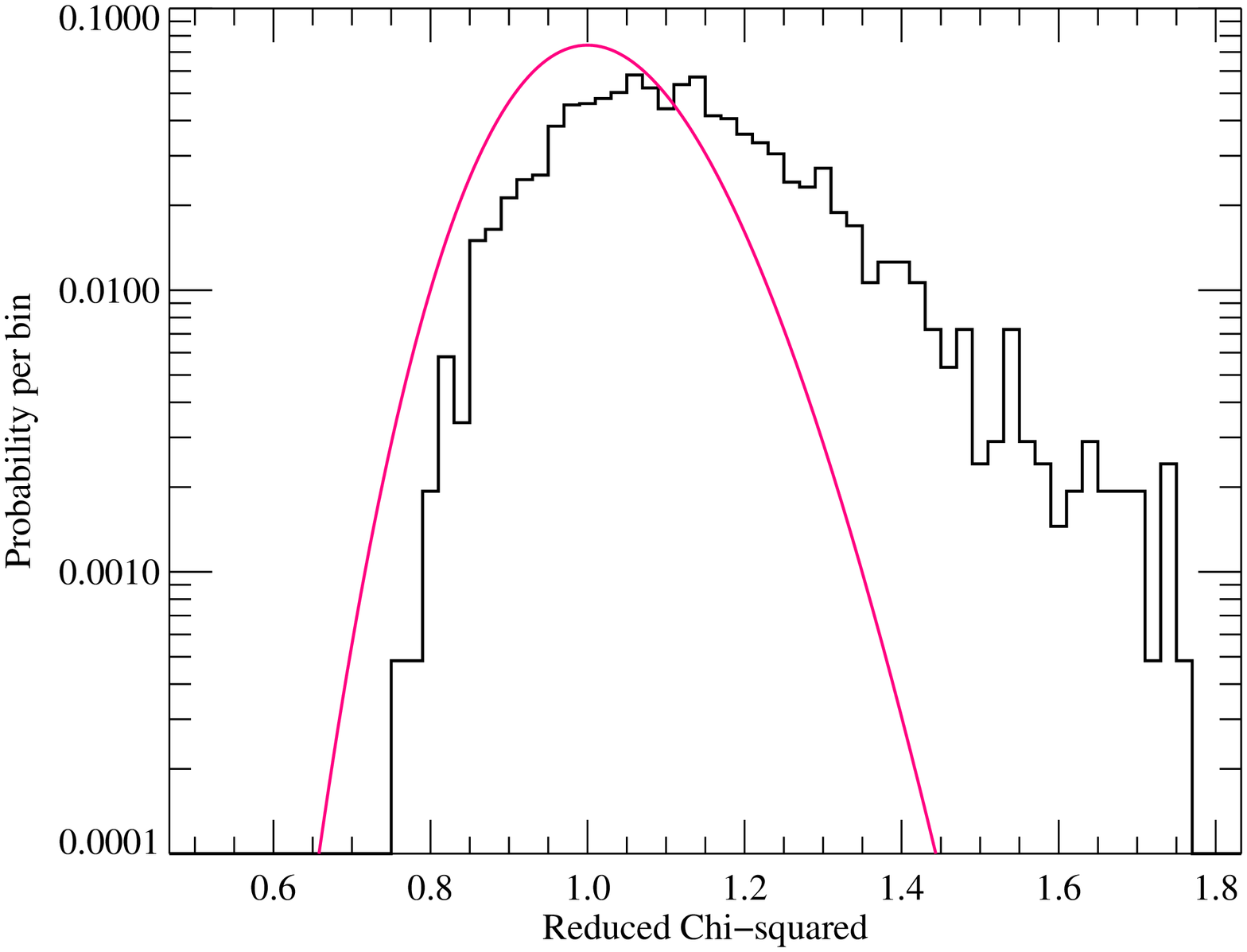}
    \end{center}
    \caption{Top: Frequency distribution (shown in black) of the reduced chi-squared values for all time intervals fitted with the \mbox{1-T} function (energy range $5.5-8.5$~keV) with expected distribution (in red) for $\nu = 75$. Bottom: Frequency distribution of reduced chi-squared values for all time intervals using the \mbox{2-T} function (energy range $1.5-8.5$~keV), with expected distribution (in red) for $\nu = 174$.}
    \label{fig:Fe1chi2prob}
\end{figure}

The occurrence distribution of $\chi^2_{\rm red}$ values is of interest as it checks whether there is a match with that expected for a chi-squared distribution with the appropriate number of degrees of freedom ($\nu = 75$: Section~\ref{companal}). Figure~\ref{fig:Fe1chi2prob} (top panel) shows the observed distribution for the \mbox{1-T} fitting function and the statistically expected distribution for $\nu = 75$. There is a very close agreement of the observed and expected distributions, showing that $1 \sigma$-normalized residuals for each energy bin are close to expected and that there are no significant instrumental effects in the $5.5-8.5$~keV energy range affecting the distributions.

% Figure 9 - Frequency distribution of chi^2 values

% Subsection 3.2
\subsection{Abundance estimates using the \mbox{2-T} function}
\label{2-T_section}

Abundance estimates of Fe, Ca, S, Si, and Ar were made using the \mbox{2-T} fitting function over the broader energy range of $1.5 - 8.5$~keV. The three free parameters describing the SAX detector energy resolution and calibration were found to vary by less than 20\% over the entire 7~yr period of the observations, indicating the stability of the SAX instrument.

Estimates of the Fe FIP bias from the \mbox{2-T} analysis are summarized in Figure~\ref{fig:Fe_vs_time} (bottom panel) and Figure~\ref{fig:Fe1vsENo} (middle panel). Figure~\ref{fig:OccvsFe1} (bottom panel) shows the frequency histograms for Fe FIP bias values from this analysis. While there is little difference in the mean value of FIP bias, $1.66 \pm 0.50$ (s.d.), or $A({\rm Fe}) = 7.72 \pm 0.11$, the amount of scatter is larger. As with the \mbox{1-T} case, there is little apparent dependence of Fe FIP bias on emission measure or temperature of the higher-temperature component. The frequency distribution of the reduced chi-squared values obtained for all analyzed time intervals in the \mbox{2-T} case deviates significantly from the expected distribution for $\nu = 174$ (Figure~\ref{fig:Fe1chi2prob}). This difference did not show up for the \mbox{1-T} fits probably because of the narrower energy range.  It most likely reflects some small instrumental effect not included in the instrument response matrix. The values of the $1 \sigma$-normalized residuals averaged over all analyzed spectra for each energy bin show that certain bins consistently have larger or smaller count fluxes at the few percent level but the origin of these small discrepancies is unknown. Since the difference is only $\sim0.1$ in the reduced chi-squared value, it is does not result in any significant deviation in the mean values of the free parameters obtained from the fits.

\begin{figure}
    \begin{center}
        \includegraphics*[angle = 90, width = 0.49\textwidth,  trim =  0 20 20 20 ] {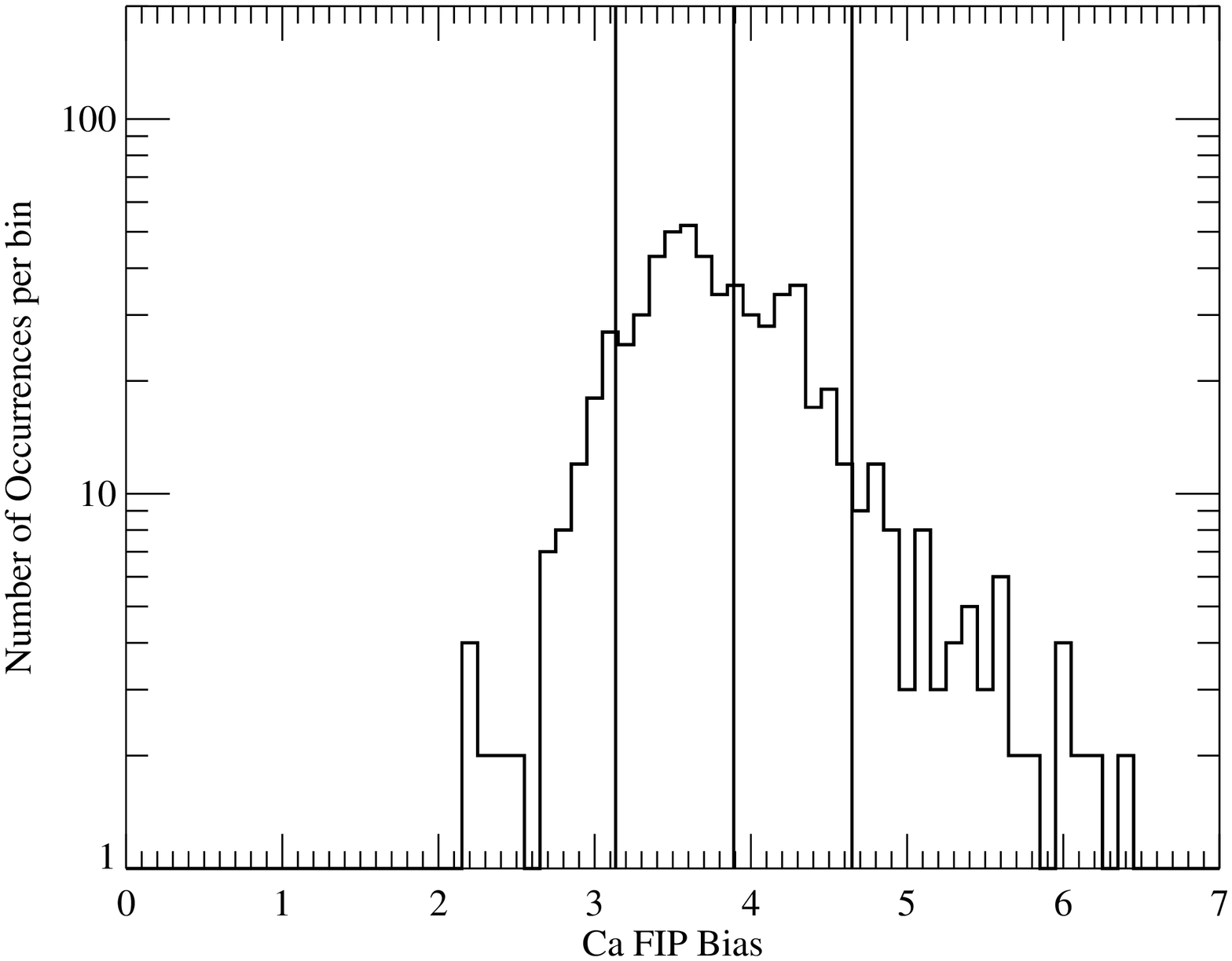}
        \includegraphics*[angle = 90, width = 0.49\textwidth,  trim =  0 20 20 20 ] {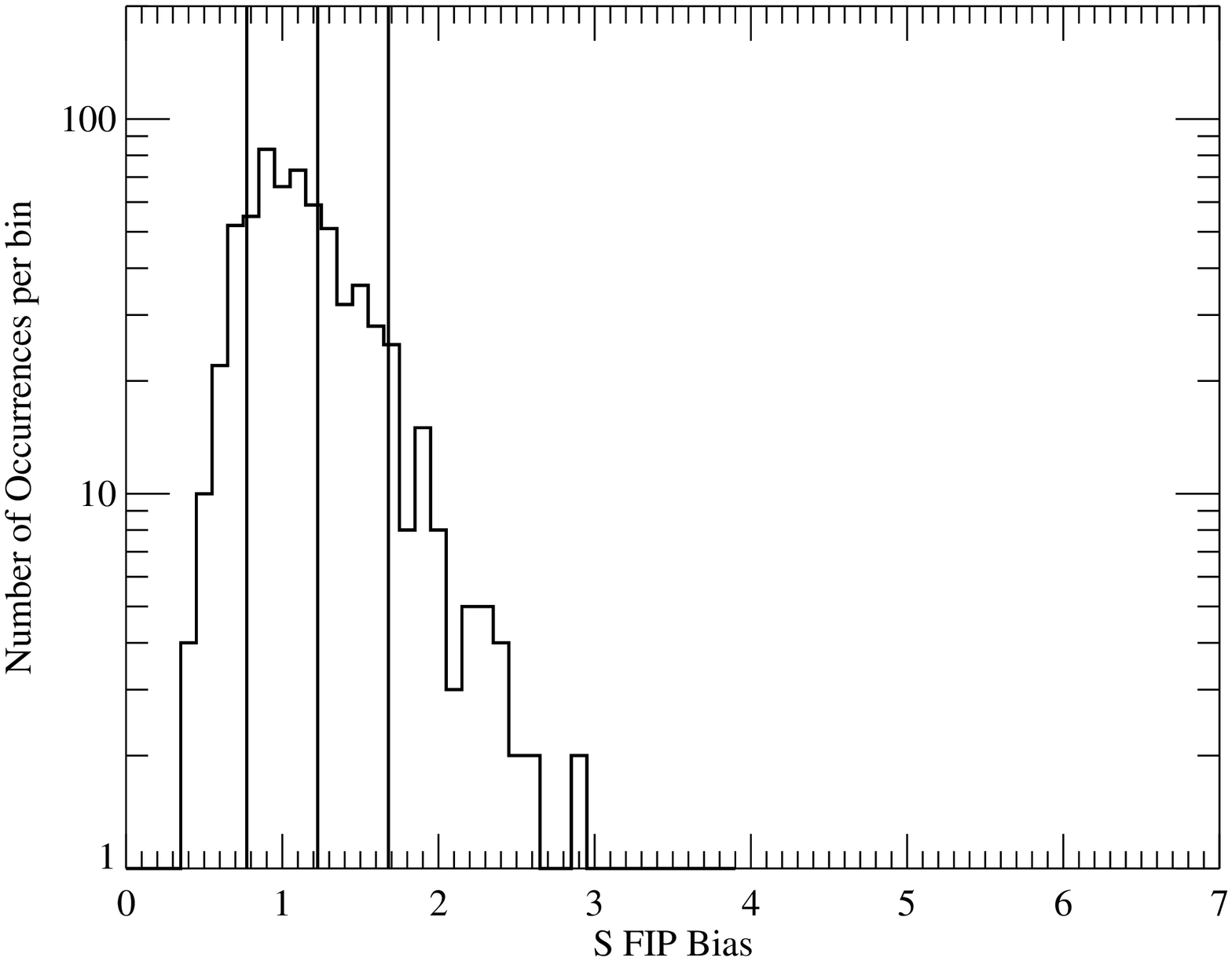}
        \includegraphics*[angle = 90, width = 0.49\textwidth,  trim = 10 20 20 20 ] {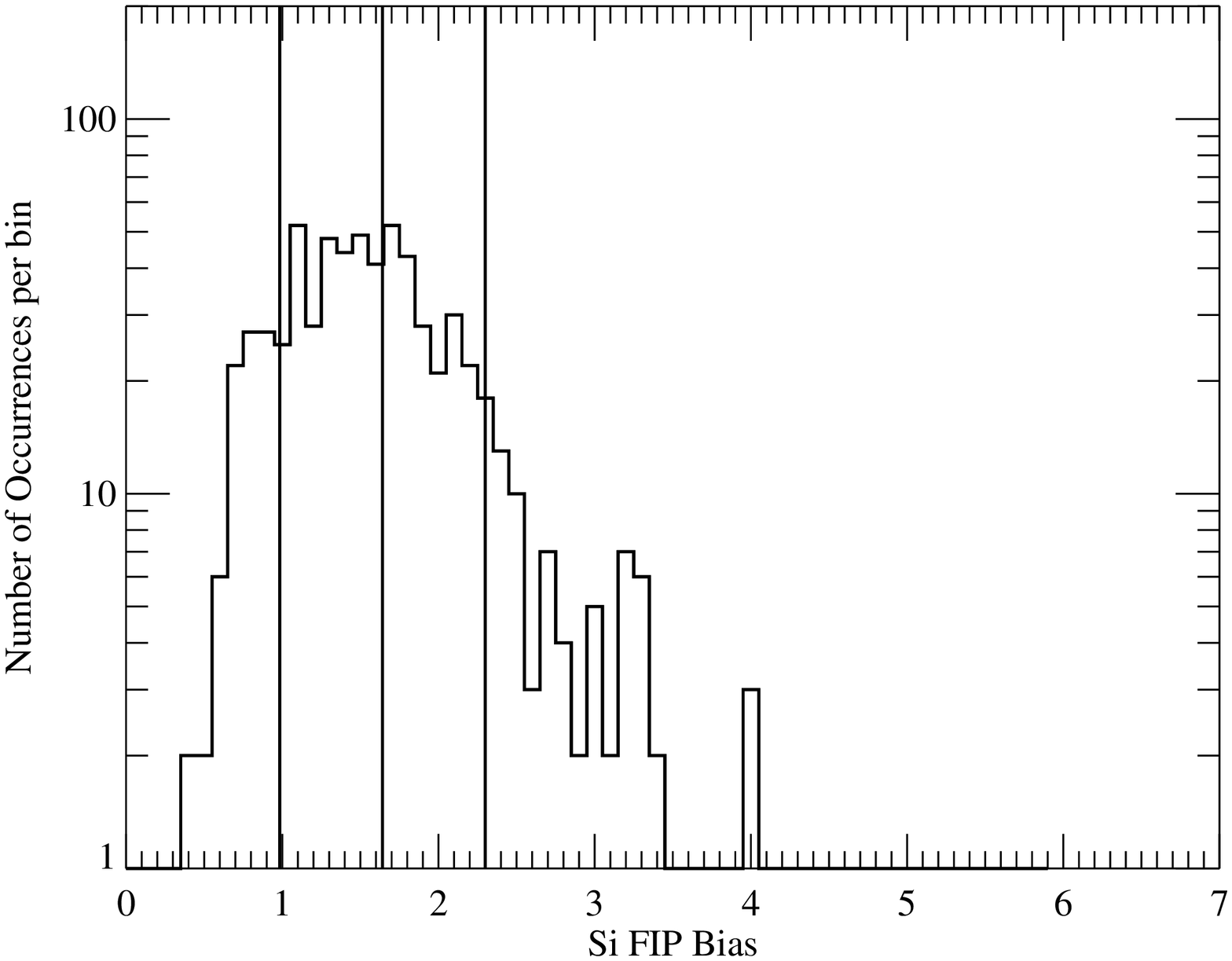}
        \includegraphics*[angle = 90, width = 0.49\textwidth,  trim = 10 20 20 20 ] {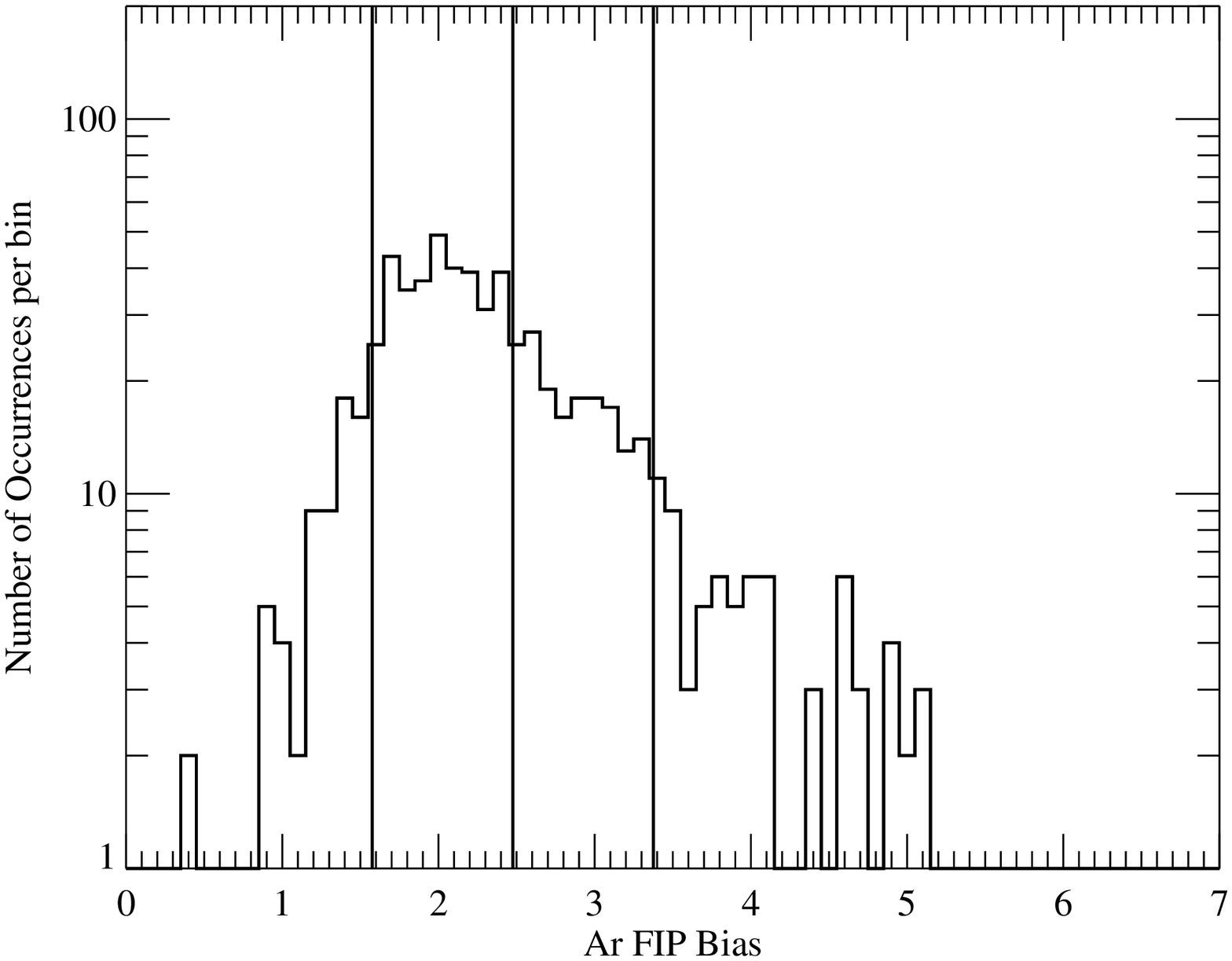}
    \end{center}
    \caption{Frequency histograms for the FIP bias values for Ca (top left), S (top right), Si (bottom left), and Ar (bottom right) determined from the \mbox{2-T} fitting function (energy range $1.5-8.5$~keV).}
    \label{fig:OccvsCaSSiAr}
\end{figure}

Abundance and frequency distribution plots were also obtained from the \mbox{2-T} function for Ca, S. Si, and Ar. Figure~\ref{fig:Fe1vsENo} (bottom panel) shows the mean Ca FIP bias for each flare plotted against flare sequence number. There is a slight but statistically significant decrease for events 238--261, as with the Fe FIP bias values shown in this plot (top two panels). Frequency histograms for Ca and also for S, Si, and Ar are given in Figure~\ref{fig:OccvsCaSSiAr}.  The mean Ca FIP bias is $3.89 \pm 0.76$ corresponding to $A({\rm Ca}) = 6.9 \pm 0.1$ with the uncertainty including any possible real variations from flare to flare.

% Figure 10 - FIP bias for Ca, S, Si, and Ar with \mbox{2-T} function

Figure~\ref{fig:OccvsCaSSiAr} (top right and bottom panels) show frequency histograms for S, Si, and Ar. As these elements do not have recognizable line complexes in SAX spectra, the FIP bias determinations are less reliable. We obtained mean values for all flares to be $1.23 \pm 0.45$ (S), $1.64 \pm 0.66$ (Si), and $2.48 \pm 0.90$ (Ar). These correspond to abundances (on a logarithmic scale with H = 12) of $A({\rm S}) = 7.2 \pm 0.2$, $A({\rm Si}) = 7.7 \pm 0.2$, $A({\rm Ar}) = 6.8 \pm 0.2$.

\begin{figure}
    \begin{center}
        \includegraphics*[angle = 90, width = 0.8\textwidth, ]
%        {abundance_vs_fip_20140529.eps}
%        {abundance_vs_fip_20141110.eps}
%       {abundance_vs_fip_20141121.eps}
        {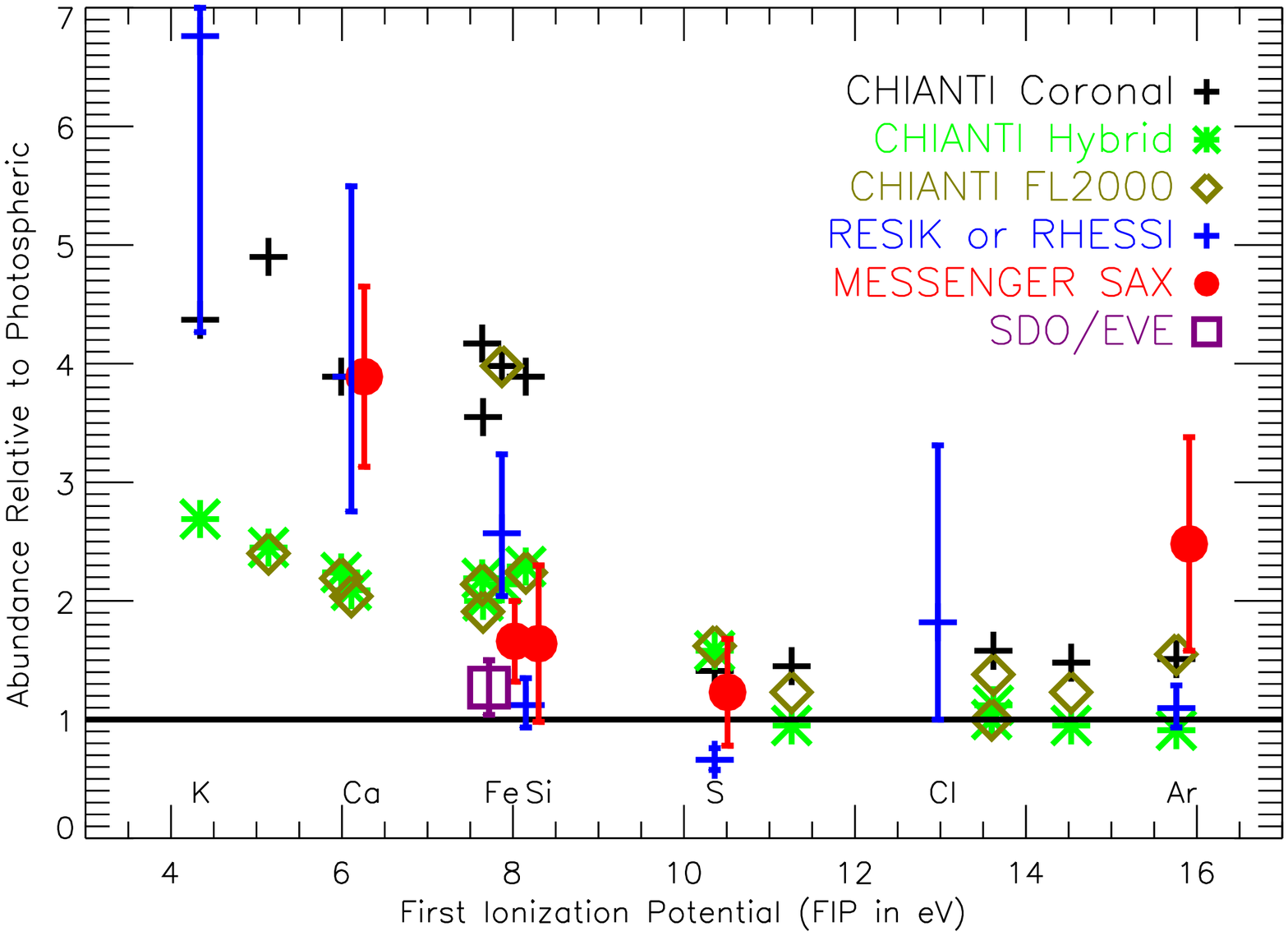}
    \end{center}
    \caption{Measured FIP bias values as a function of FIP. Red circles: current results for Fe, Ca, Si, S, and Ar (FIP bias values moved to the right by 0.15~eV for clarity).  The error bars show the $\pm 1\sigma$ values obtained from the scatter of the measured values shown in Figure~\ref{fig:OccvsFe1} for Fe and Figure~\ref{fig:OccvsCaSSiAr} for other elements. Also shown are the following published FIP bias values: black plus signs: {\sc chianti} ``coronal'' abundance set \citep{asp09}; green asterisks: {\sc chianti} ``hybrid'' abundance set \citep{flu99}; purple square: Fe FIP bias from SDO/EVE given by \cite{war14} (moved to the left by 0.15 eV for clarity); blue plus signs with $\pm1\sigma$ error bars: RESIK K, Ca, Si, S, Ar values given by \cite{bsyl14} and {\em RHESSI} Fe value given by \cite{phi12}.}
    \label{fig:FIP_bias_full}
\end{figure}

The mean FIP bias values and corresponding element abundances determined from this analysis are given in Table~2.  The FIP bias is plotted against the FIP for each element in Figure~\ref{fig:FIP_bias_full}.

% Table 2
\begin{deluxetable}{crc ccccc cc}
\tabletypesize{\scriptsize}
%\rotate
\tablecaption{FIP Biases from SAX spectra for 526 flares (2007 May~28--2013 August~19).}

\tablewidth{0pt}

%%& \multicolumn{4}{c}{Element Abundances A(H) = 12} & \multicolumn{6}{c}{FIP Bias - Ratio to Asplund (2009) Photospheric Abundances}\\
\tablehead{
& \colhead{FIP} & \colhead{Photo-$^a$} & \multicolumn{2}{c}{{\sc chianti}} & \colhead{RESIK} & \colhead{SAX} &  \colhead{SAX} & \colhead{XSM$^b$} & \colhead{SDO/EVE$^c$}  \\
 & \colhead{(eV)} & \colhead{sphere} & \colhead{Corona} & \colhead{Hybrid} & \colhead{RHESSI} & \colhead{2007$^d$} & \colhead{2007-13} &   &  \\
         }
\startdata
%Si & 8.15 & 7.51 & 8.10 & 7.56 & 3.89 & 2.29 & 1.12 &  2.15 & 1.49$\pm$0.48 &  \\
%S & 10.36 & 7.12 & 7.27 & 6.94 & 1.41 & 1.58 & 0.66 &  0.39 & 1.11$\pm$0.34 &  \\
%Ar & 15.76 & 6.40 & 6.58 & 6.44 & 1.51 & 0.91 & 1.10 & 1.48 & 2.22$\pm$0.90 &   \\
%K & 4.34 & 5.03 & 5.67 & 5.86 & 4.37 &  & 6.76 &  &  &   \\
%Ca & 6.11 & 6.34 & 6.93 & 6.93 & 3.89 & 2.09 & 3.89 &  3.59 & 3.52$\pm$0.69 &    \\
%Fe & 7.87 & 7.50 & 8.10 & 7.91 & 3.98 & 2.14 & 2.57$\pm$0.60 & 1.70 & 1.66$\pm$0.50 & 1.27$\pm$0.22 \\
Fe & 7.87 & 7.50 & 3.98 & 2.14 & 2.57 & 1.70 & 1.66$\pm$0.34 & 3.09$\pm$0.80 & 1.27$\pm$0.22 \\
Ca & 6.11 & 6.34 & 3.89 & 2.09 & 3.89 & 3.59 & 3.89$\pm$0.76 & 1.82$\pm$1.06 &  \\
S & 10.36 & 7.12 & 1.41 & 1.58 & 1.10 & 0.78 & 1.23$\pm$0.45 & 1.51$\pm$1.51 &  \\
Si & 8.15 & 7.51 & 3.89 & 2.29 & 2.34 & 1.09 & 1.64$\pm$0.66 & 2.63$\pm$1.64 &  \\
Ar & 15.76 & 6.40 & 1.51 & 0.91 & 1.10 & 1.48 & 2.48$\pm$0.90 &  &  \\
K & 4.34 & 5.03 & 4.37 & 2.69 & 6.76 &  &  &  &  \\
\enddata

\tablenotetext{a} {Absolute abundances from \cite{asp09} on logarithmic scale with A(H) = 12.}
\tablenotetext{b} {FIP bias values from \cite{nar14}.}
\tablenotetext{c} {FIP bias values from \cite{war14}.}
\tablenotetext{d} {FIP bias values from Table~\ref{tab:JuneflareFIP&FIPbias} for the \mbox{2-T} fits to the 2007 June 1 flare.}
\label{tab:table2}
\end{deluxetable}

% Figure 11 - Element abundances (FIP bias) vs. FIP, all elements

%%%%%%%%%%%%%%%%%%%%%%%%%%%%%%%%%%%%%%%%%%%%%%%%%%%%%%%%%%%%%%%%%%%%%%%%%%%%%%%%%%%%%%%%%%%%%%%%%%%%%%%%%%%%%%%%%%%%%%%

% Section 4
\section{DISCUSSION AND CONCLUSIONS}
\label{discussion}

The FIP bias results from our analysis of {\em MESSENGER} SAX solar flare data cover the period 2007--2013, when the spacecraft went from the vicinity of the Earth to orbit about Mercury. They are summarized in Table \ref{tab:table2}.

Other investigators have estimated flare abundances of the elements considered here by various methods and so it is of interest to compare their results with ours. {\em RHESSI} is a broad-band X-ray spectrometer like SAX covering the energy range above $\sim$5~keV with a slightly poorer spectral resolution of $\sim 1$~keV but still adequate to resolve the Fe-line complex at 6.7 keV. Analysis by \cite{phi12} of the decay phase of 20 strong flares ({\em GOES} class up to X8) seen with {\em RHESSI} gave $A({\rm Fe}) = 7.91 \pm 0.10$ or a FIP bias of $2.6 \pm 0.6$.  This is higher than the result obtained here though just within the statistical uncertainties of each determination. The X-ray Solar Monitor (XSM) on the Indian lunar space mission {\em Chandrayaan-1} observed solar $1.8-20$~keV X-ray emission to complement the X-ray fluorescence instrument on board designed to investigate the lunar surface composition. XSM had a spectral resolution of 200~eV, an improvement over SAX and enough to resolve spectral features of all the elements reported here. The mission only worked for nine months during a solar-quiet period and no strong flares were observed. However, a C2.8 flare in 2009 was analyzed by \cite{nar14}, and they reported a weighted average $A({\rm Fe}) = 7.99 \pm 0.09$ corresponding to a FIP bias of $3.1 \pm 0.7$, which is similar to the \cite{phi12} {\em RHESSI} result but higher than the SAX value obtained here. For several smaller flares, \cite{nar14} reported $A({\rm Ca}) = 6.6 \pm 0.2$ and $A({\rm S}) = 7.3 \pm 0.3$, consistent with our abundance estimates.

Broad-band X-ray spectrometers, in general, have an inferior spectral resolution to that of crystal spectrometers flown on numerous spacecraft since the 1960s.  However, they have the advantage that the continuum is unambiguously observed whereas many crystal spectrometers suffered from a large background formed by the fluorescence of the crystal material by solar X-rays. This difficulty has been overcome in some cases by using crystals made with material having low atomic number, as with the graphite crystals in the {\em OSO-8} spectrometer. Solar flare data from this instrument were analyzed by \cite{vec81} giving $A({\rm Si}) = 7.7^{+0.2}_{-0.3}$, $A({\rm S}) = 6.9^{+0.1}_{-0.3}$, $A({\rm Ar}) = 6.4^{+0.2}_{-0.3}$, and $A({\rm Ca}) = 6.5^{+0.1}_{-0.2}$. While the Si abundance is very close to our value, S, Ar, and Ca are all lower than ours and just outside the range indicated by the error bars. The fluorescence background was reduced to zero or to an accurately measured amount in the RESIK crystal spectrometer.  A differential emission measure analysis of RESIK observations of a large flare gave $A({\rm Si}) = 7.56 \pm 0.08$, which is consistent with our result, and $A({\rm S}) = 6.94 \pm 0.06$ which, like the \cite{vec81} result, is lower than our result. Earlier analysis of the Ca abundance during many flares observed by the {\em Solar Maximum Mission} Bent Crystal Spectrometer by \cite{syl98} showed that there was a factor-of-three variation from flare to flare, with a mean value $A({\rm Ca}) = 6.76 \pm 0.10$, consistent with our Ca abundance and its uncertainty. Thus, in the light of this result, the apparent flare-to-flare abundance variations found here (e.g., the slight decrease for flares 238--261) may  be real.

\cite{war14} used the fluxes of lines due to Fe ionization stages from \ion{Fe}{15} to \ion{Fe}{24} obtained from the Solar Dynamics Observatory Extreme ultraviolet Variability Experiment (EVE) (wavelength range: 60-200~\AA). His result for the FIP bias of Fe using a DEM analysis was $1.17 \pm 0.22$, corresponding to $A({\rm Fe}) = 7.57$, lower than obtained here or by \cite{phi12}. The form for the temperature dependence of the DEM was taken to be the sum of Gaussian functions fixed at particular values of log~$T$, an artificial assumption which is far from the near-bimodal functions derived by other authors for solar X-ray flares \citep[e.g.,~][]{mct99,bsyl14}. Also, there is little constraint on the high-temperature part of the DEM function which may lead to some uncertainty in its determination, while the crowded nature of emission lines in the (e.g.,) $100-120$~\AA\ region is only just resolved by the $\sim 1$~\AA\ spectral resolution of EVE.

Element enhancements in closed coronal loops by a FIP effect have been theoretically modeled by \cite{lam04}. The idea is that the more easily ionized low-FIP elements are preferentially accelerated into the corona by a ponderomotive force associated with Alfv\'{e}n waves that travel in the loop down to the chromospheric footpoints. The latest evolution of this theory also includes slow-mode waves \citep{lam12}. The particular case in these calculations most appropriate to solar flares indicates enhancements by factors of 2--3 (relative to oxygen) for Ca and Fe and more modest enhancements for S and Si and no enhancement for the high-FIP element Ar. Apart from Fe for which the predicted value is larger than obtained here, our results have some resemblance to the predictions of one of the model calculations with small wave amplitudes. However, the 100,000~km loop length in the calculations is much larger than typical X-ray flare dimensions as revealed by {\em RHESSI} images, so that a more detailed comparison with realistic loop dimensions and other parameters is clearly desirable.

The small enhancements over photospheric abundance indicated in this work for the relatively high-FIP elements S and Ar are consistent with the original ideas expressed by \cite{fel92} and \cite{fel00} on the FIP effect for coronal plasmas but our low enhancements for Si and Fe (FIP values $\sim 8$~eV) are not expected. The much higher enhancement that we find for Ca (FIP $= 6.1$~eV) still indicates the presence of a FIP effect for solar flare plasma, as does the very high enhancement factor found from RESIK spectra of $5.8 \pm 1.6$ \citep{syl10a}) for K with its very low FIP of 4.3~eV. This would seem to point to a boundary between low-FIP and high-FIP elements that is significantly less than the 10~eV that is normally assumed, perhaps about 7~eV.  This would imply that the fractionation takes place at a lower temperature than has been conceived of in the past. Results for a white-light flare \citep{oli12} point to a similar low altitude in the solar atmosphere of white-light emission and a hard X-ray (30--40~keV) source as observed by {\em RHESSI} for a near-limb flare, around 200--300~km above the photosphere, below the temperature minimum and considerably below the altitude expected from the standard thick-target model for hard X-ray emission. If this applies to flares generally, the origin of the flare plasma may be at lower chromospheric altitudes than had been previously thought. Likewise, the deviations from photospheric abundances in flare plasmas may arise from a fractionation process, perhaps like that described by \cite{lam04} but operating at lower temperatures than has previously been considered. The apparently time-varying Ca and Fe abundances from flare to flare (as indicated for both Ca and Fe in this work and Ca from \cite{syl98}) should also be considered in any theoretical modeling of the FIP effect.

\acknowledgments

We acknowledge the {\em MESSENGER} XRS project team for making all data promptly and publicly available for independent analysis. {\sc chianti} is a collaborative project involving the US Naval Research Laboratory, the Universities of Florence (Italy) and Cambridge (UK), and George Mason University (USA).

{\em Facilities:} \facility{RHESSI}, \facility{GOES}

\bibliographystyle{apj}	%% AASTeX

\bibliography{abundances}

\end{document}